\documentclass[fleqn,usenatbib]{mnras}
\usepackage{newtxtext,newtxmath}
\usepackage[T1]{fontenc}
\DeclareRobustCommand{\VAN}[3]{#2}
\let\VANthebibliography\thebibliography
\def\thebibliography{\DeclareRobustCommand{\VAN}[3]{##3}\VANthebibliography}
\usepackage{graphicx}	
\usepackage{amsmath}	
\usepackage{orcidlink}
\usepackage{caption}
\usepackage{adjustbox}
\usepackage{booktabs} 
\usepackage{subcaption} 

\title[Two Distinct Epochs of FRB 20240114A]{Signatures of Two Distinct Epochs of FRB 20240114A from January to August 2024 Based on its Energy and Waiting Time Analysis}

\author[Xiao Li et al.]{
Xiao Li\orcidlink{0009-0001-5037-3855}$^{1}$, Ying Gu\orcidlink{0009-0005-6943-7803}$^{1}$ and 
En-Wei Liang\orcidlink{0000-0002-7044-733X}$^{1}$\thanks{lew@gxu.edu.cn} 
\\
$^{1}$Guangxi Key Laboratory for Relativistic Astrophysics, School of Physical Science and Technology, Guangxi University, Nanning 530004, China\\}
\date{Accepted XXX. Received YYY; in original form ZZZ}
\pubyear{\the\year{}}
\begin{document}
\label{firstpage}
\pagerange{\pageref{firstpage}--\pageref{lastpage}}
\maketitle

\begin{abstract}
A comprehensive analysis of the energy and waiting time distributions of the bursts from FRB 20240114A detected by the Five-hundred-meter Aperture Spherical Radio Telescope between 28 January and 29 August 2024 is presented. For the full sample, its energy distribution cannot be fitted with the simple power-law (SPL), bent power-law (BPL), thresholded power-law (TPL) or Band function models, and its waiting time distribution excluding intervals shorter than 0.5 s cannot be fitted with the Poisson or Weibull models. 
Nevertheless, for the subsamples with more than 50 bursts in single-day observations, their energy distributions can be fitted with the BPL or TPL models, and their waiting time distributions are better described by a Weibull model. 
It is noted that the best-fitting BPL parameter $\beta$ is approximately invariant within the epochs before and after 21 March 2024, with an average of $\bar \beta_b = 1.006 \pm 0.074$ and $\bar \beta_a = 1.236 \pm 0.183$ (one standard deviation), respectively. Most subsamples from the later epoch have a smaller burst rate parameter $r$ in the Weibull model than those from the earlier epoch. The majority of bursts with $E>10^{39}$ erg occurred in the earlier epoch. 
The energy distributions in the high-energy range ($> 6\times10^{37}$ erg) differ significantly between the two epochs, and power-law fits to $dN/dE$ yield indices of $-1.97_{-0.02}^{+0.02}$ and $-2.34_{-0.06}^{+0.06}$, respectively.
The median of the waiting time distribution of the later epoch is larger than that in the earlier epoch. These results suggest that the two epochs may be dominated by different types of bursts, possibly attributed to changes in the physical properties of the emission region.

\end{abstract}

\section{Introduction} 
Fast radio bursts (FRBs) are bright, short-duration radio transients that occur at cosmological distances, whose physical origin remains elusive \citep{2007Sci...318..777L,2019ARA&A..57..417C,2022A&ARv..30....2P,2023RvMP...95c5005Z}. The discovery of an association between the magnetar SGR 1935+2154 and FRB 20200428 suggests that at least some FRBs may originate from magnetars \citep{2020Natur.587...59B,2020Natur.587...54C}. To date, more than 800 FRBs have been detected \citep{2016PASA...33...45P,2021ApJS..257...59C,2024ApJ...969..145C}\footnote{\url{https://blinkverse.zero2x.org/overview}}. On the basis of their burst activity patterns, FRBs are observationally classified into repeaters (repeating FRBs) and non-repeaters (apparently one-off FRBs). Whether all FRBs are repeaters remains under debate \citep{2019MNRAS.484.5500C,2021ApJ...906L...5A,2022ApJ...926..206Z,2024ApJ...975...75L,2025ApJ...993...37B}.

Although most known FRBs appear to be non-repeaters, it has been reported that about 8\% of them are repeaters \citep{2023ApJ...947...83C}. About 7\% of these repeaters have demonstrated hyperactivity, such as FRB 20121102A, FRB 20180916B, FRB 20190520B, FRB 20201124A, and FRB 20220912A, which exhibit very high burst rates and have accumulated large amounts of observational data from multiple telescopes (e.g., \citealt{2016Natur.531..202S,2021Natur.598..267L,2021ApJ...922..115A,2022MNRAS.515.3577H,2023MNRAS.519..666J,2020Natur.582..351C,2024arXiv240920307B,2022Natur.606..873N,2022Natur.609..685X,2022RAA....22l4001Z,2024NatAs...8..337K,2025A&A...696A.194B,2023ApJ...955..142Z,2024MNRAS.534.3331K,2024MNRAS.529.1814H,2024ApJ...974..296F,2025arXiv250715790W}).The bursts of some hyperactive repeaters are predominantly confined to distinct active episodes that typically last from several days to several months \citep{2021Natur.598..267L,2023MNRAS.519..666J,2025arXiv250715790W,2022Natur.609..685X,2022RAA....22l4001Z,2024NatAs...8..337K}. While FRB 20180916B and FRB 20121102A exhibit periodic activity \citep{2020Natur.582..351C,2020MNRAS.495.3551R,2021MNRAS.500..448C}, long-term monitoring has shown that bursts do not occur in every predicted active window \citep{2024MNRAS.527.9872G,2025A&A...693A..40B,2025A&A...704A..25G}. 

The statistical properties of large datasets of hyperactive repeaters, such as the distributions of energy and waiting time (defined as the detected time interval between two adjacent bursts), provide important insights into their origins and physical mechanisms \citep{2018MNRAS.475.5109O,2019ApJ...882..108W,2021MNRAS.500..448C,2021ApJ...920L..23Z}. However, observational evidence shows that the distributions of energy and waiting time among different active episodes of the same hyperactive repeater vary significantly. For example, FRB 20121102A exhibited a bimodal energy distribution during its active episode in 2019, whereas this feature disappeared in the 2022 and 2023 episodes, while its waiting time distributions remained bimodal but with the second peak shifting to shorter timescales \citep{2021Natur.598..267L,2025arXiv250715790W}.
Similarly, the cumulative energy distribution of FRB 20201124A changes from a broken power-law to a Band function between different active episodes, with the second peak of the bimodal waiting time distribution also moving to a shorter timescale \citep{2022Natur.609..685X,2022RAA....22l4002Z}. In addition, \cite{2023MNRAS.523.5430S} noted that even within a single active episode, the energy and waiting time distributions of repeaters evolve over time. This temporal evolution of energy and waiting time statistics suggests that different burst types may dominate at different epochs.

FRB 20240114A is a repeater first discovered by the Canadian Hydrogen Intensity Mapping Experiment (CHIME), with a dispersion measure (DM) of $\sim 527.7 \rm \;pc\; cm^{-3}$ and a rotation measure (RM) of $\sim+320 \rm \; rad\;cm^{-2}$ \citep{2024ATel16420....1S,2025arXiv250513297S}.
It was localized to a star-forming galaxy at a redshift of $z=0.1306\pm0.0002$ behind a foreground galaxy cluster \citep{2024ATel16613....1B,2024MNRAS.533.3174T,2025arXiv250611915B,2025ApJ...980L..24C}. 
Many radio telescopes have conducted follow-up observations of FRB 20240114A, including Parkes/Murriyang telescope \citep{2024ATel16430....1U}, the Westerbork RT1 25-m telescope \citep{2024ATel16432....1O}, Five-hundred-meter Aperture Spherical Radio Telescope (FAST) \citep{2024ATel16433....1Z,2024ATel16505....1Z,2025arXiv250714707Z}, MeerKAT \citep{2024MNRAS.533.3174T}, the upgraded Giant Metrewave Radio Telescope (uGMRT) \citep{2025ApJ...989...15P,2024ApJ...977..177K}, European Very Long Baseline Interferometry (EVN) PRECISE \citep{2024ATel16542....1S}, the Nancay Radio Telescope (NRT) \citep{2024ATel16597....1H}, the Effelsberg 100-m Telescope \citep{2024ATel16620....1L,2025arXiv251008367L,2025A&A...695L..10E}, the Northern Cross (NC) radio telescope \citep{2024ATel16434....1P}, the Robert C. Byrd Green Bank Telescope (GBT) \citep{2025ApJS..278...49X}, the Shanghai Tianma Radio Telescope (TMRT) \citep{2025arXiv250815615W}, and the Kunming 40-Meter Radio Telescope (KM40M) \citep{2025RAA....25h5009H}. 
Tens of thousands of bursts from FRB 20240114A have been detected by the aforementioned radio telescopes in the frequency range of 0.5-6.0 GHz, with FAST contributing the largest observational sample. 

In this paper, based on a comprehensive analysis of the energy and waiting time distributions of FRB 20240114A detected by FAST, we investigate whether its burst signatures changed over a seven-month observing period. 
This paper is structured as follows: Section \ref{sec:2} details the data and models. Section \ref{sec3} presents the results of the analysis. 
Discussions and conclusions are presented in Sections \ref{sec4} and \ref{sec5}, respectively.

\begin{figure*}
\centering
\includegraphics[width=0.45\textwidth]{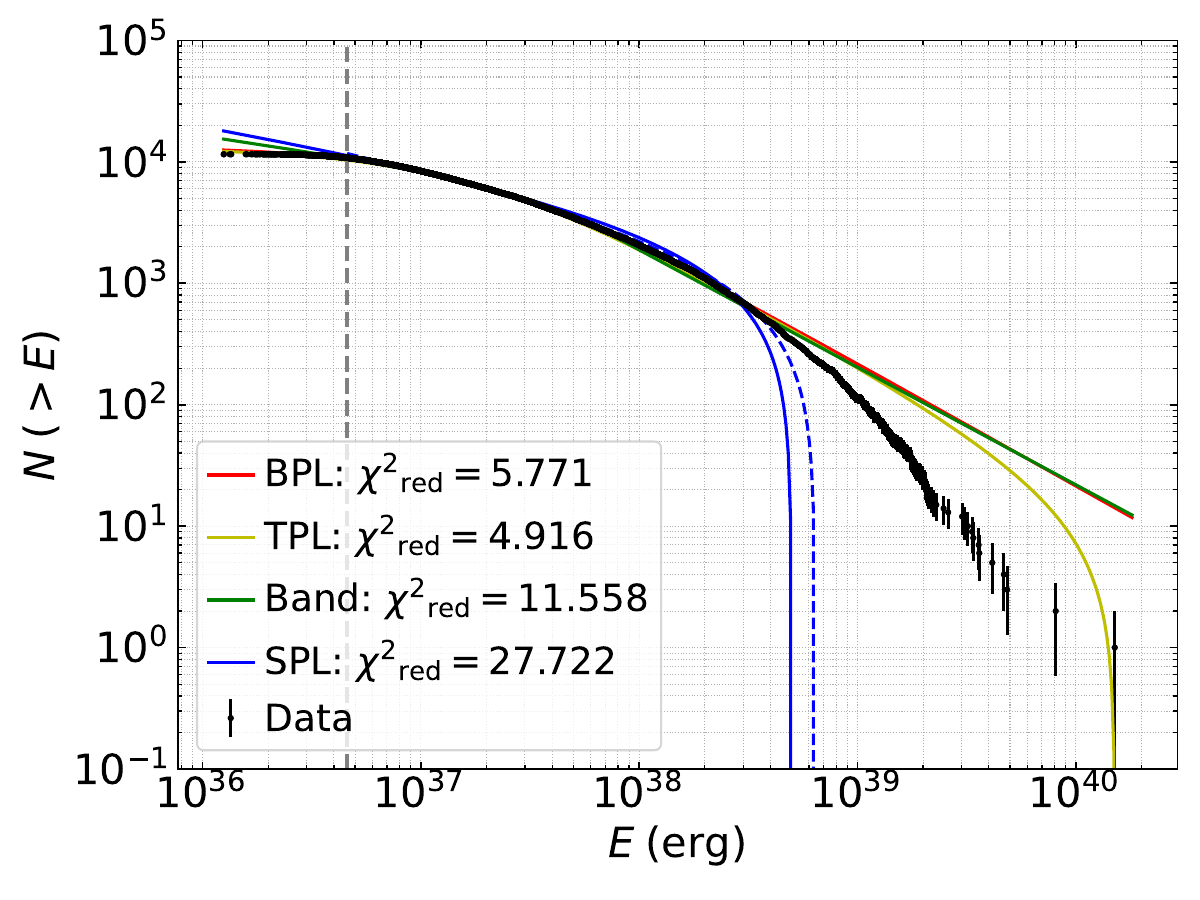}
\hspace{0.03\textwidth} 
\includegraphics[width=0.45\textwidth]{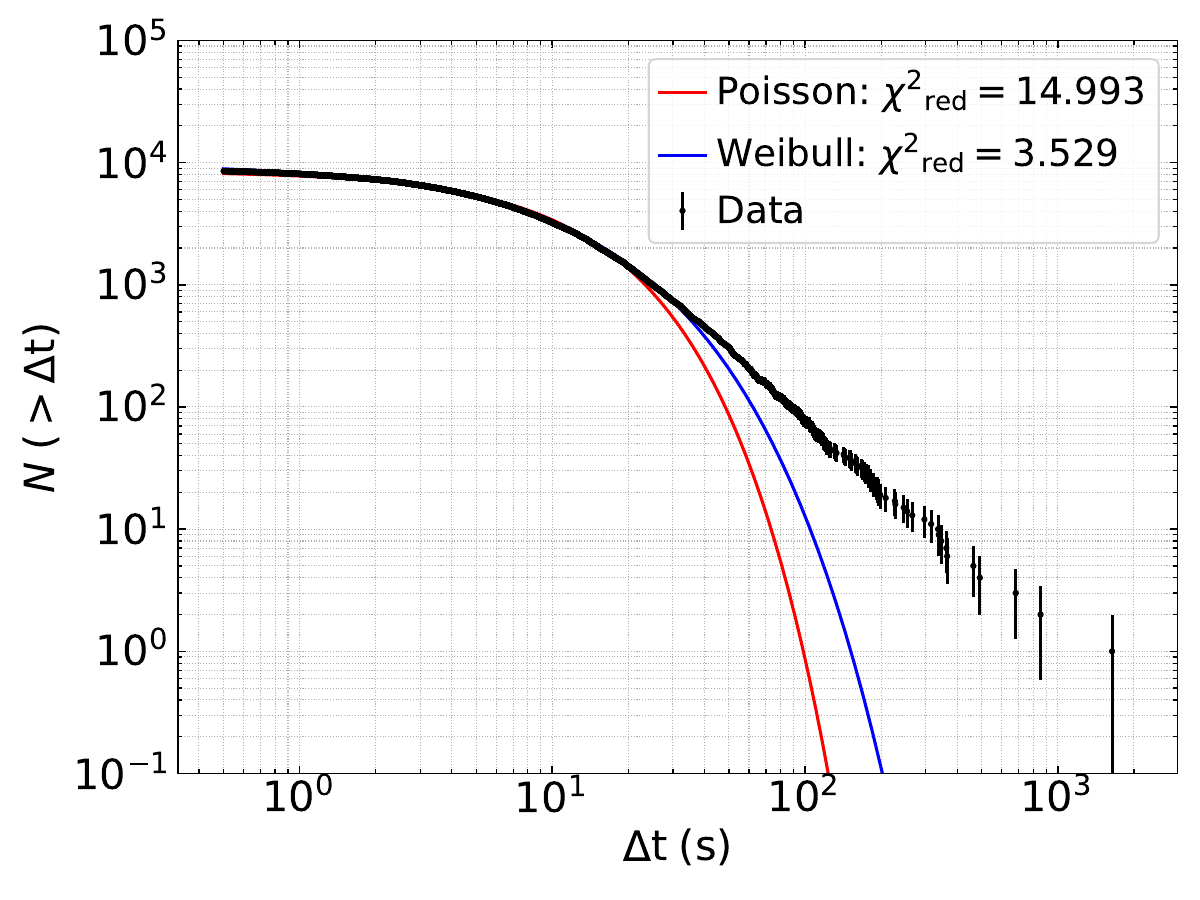}
\includegraphics[width=0.45\textwidth]{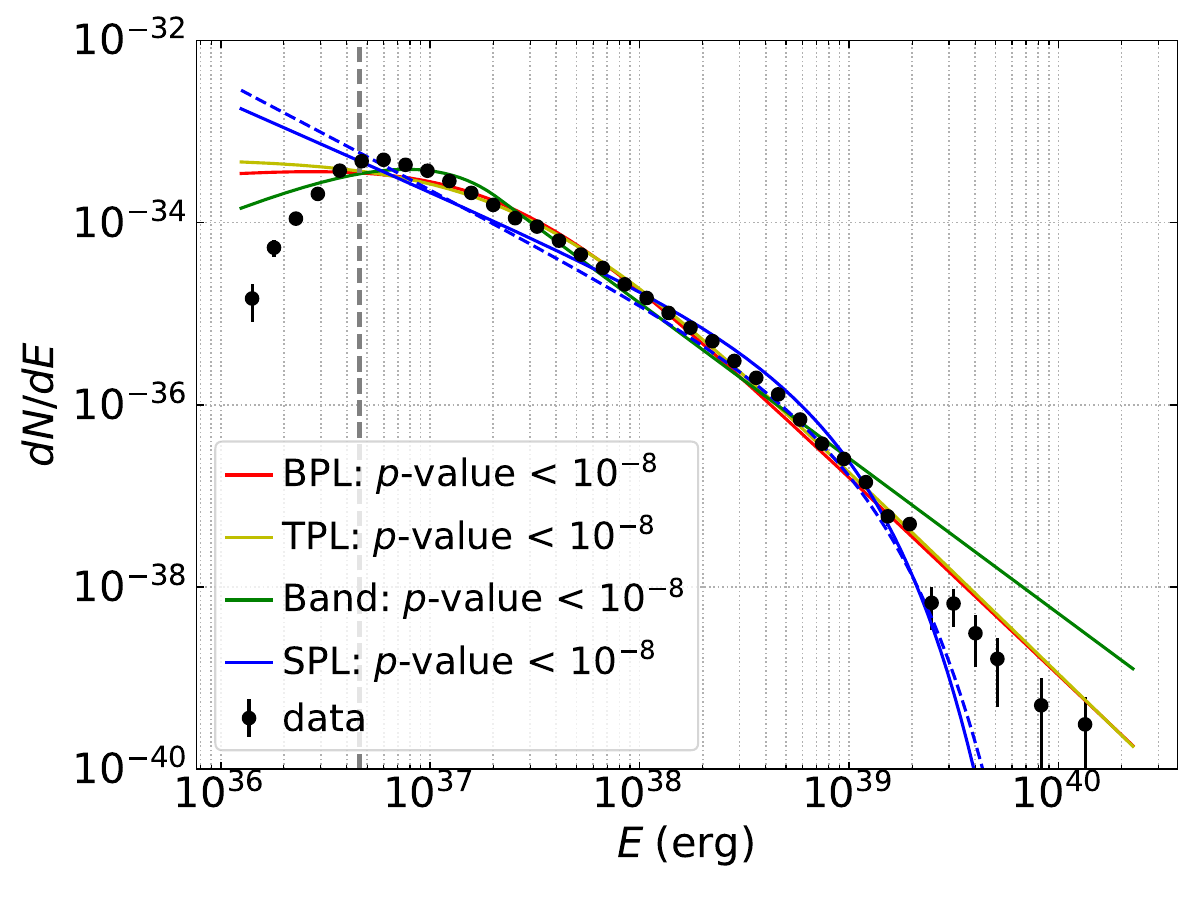}
\hspace{0.03\textwidth} 
\includegraphics[width=0.45\textwidth]{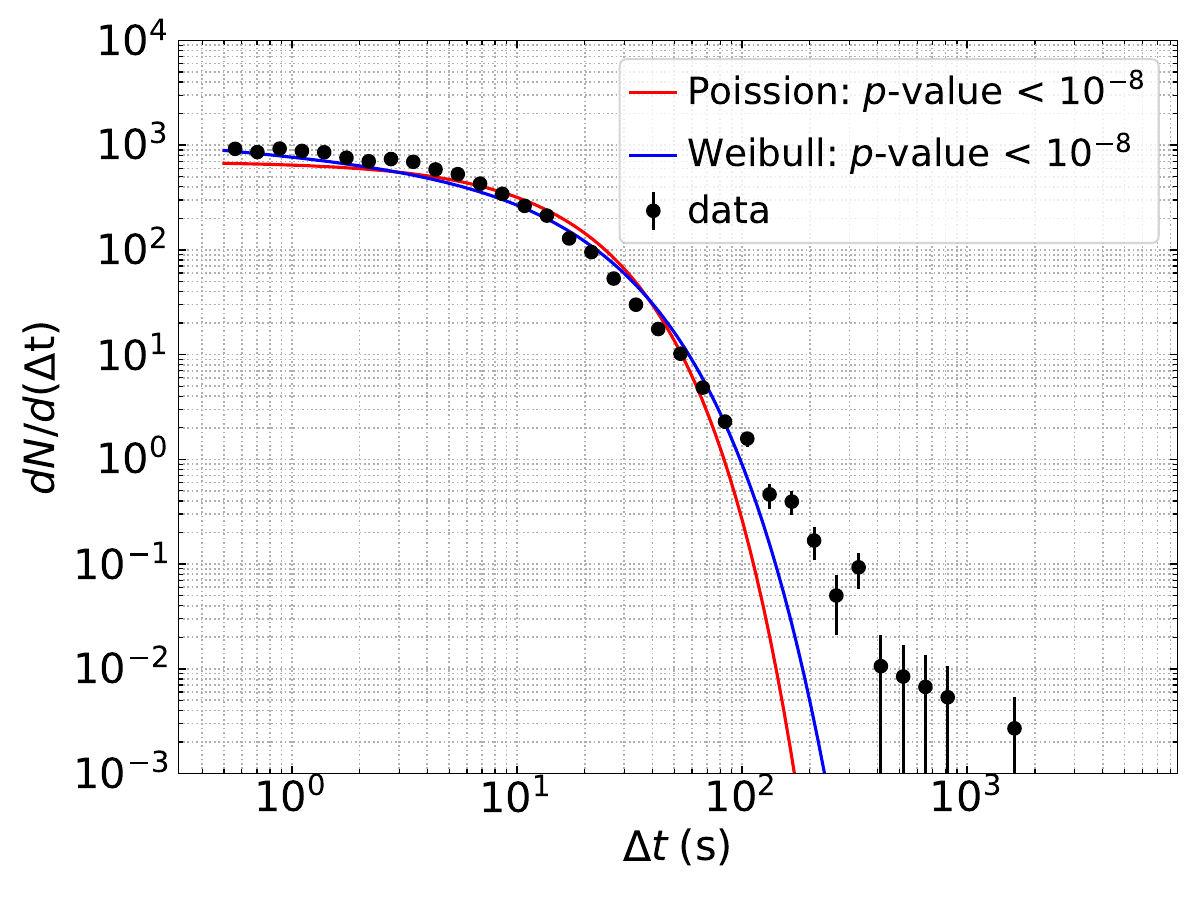}
\caption{{\em Left panels---} The cumulative and differential distributions of energy for the bursts of FRB 20240114A, along with the best-fitting SPL ({\em blue}), BPL ({\em red}), TPL ({\em green}), and Band ({\em yellow}) models.
The dashed gray line indicates the 90\% detection completeness threshold of FAST, and the blue dashed line shows the best-fitting SPL model obtained using only data above this threshold.  
An exponential cutoff is introduced in the differential SPL model.
{\em Right panels---} 
Same as the left panels, but for the waiting time distributions, along with best-fitting Poisson ({\em red}) and Weibull ({\em blue}) models. The differential energy and waiting time distributions are derived from the unbinned maximum-likelihood analysis (see details in Appendix \ref{app}).
}\label{Fig1}
\end{figure*}

\begin{table*}
\centering
\caption{The best-fitting parameters of different models for the CDFs of energy and waiting time of FRB 20240114A.}
\begin{tabular*}{\textwidth}{@{\extracolsep{\fill}}lllll} 
\hline\hline 
\multicolumn{5}{c}{energy}\\
\hline
SPL & & $\alpha=0.276\pm0.001$  & $E_c=(4.990\pm0.025)\times{\rm 10^{38}erg}$ & $\chi^2_{\rm red}=27.722$\\
BPL & & $\beta=1.011\pm0.001$  & $E_b=(1.720\pm0.005)\times{\rm 10^{37}erg}$ & $\chi^2_{\rm red}=5.771$\\
TPL & & $\gamma=2.012\pm0.002$ & $E_0=(1.740\pm0.007)\times{\rm 10^{37}erg}$ & $\chi^2_{\rm red}=4.916$\\
Band & $\hat\alpha=-0.258\pm0.001$ & $\hat\beta=-0.967\pm0.003$ & $\hat{E_{ c}}=(9.611\pm 0.059)\times{\rm 10^{37}erg}$ & $\chi^2_{\rm red}=11.558$\\
\hline
\multicolumn{5}{c}{waiting time}\\
\hline
Poisson & & & $\lambda=(0.919\pm0.001)\times{\rm 10^{-1}s^{-1}}$ & $\chi^2_{\rm red}=14.993$\\
Weibull & & $k=0.777\pm0.001$ & $r=(0.989\pm0.001) \times{\rm 10^{-1}s^{-1}}$ & $\chi^2_{\rm red}=3.529$
\\
\hline 
\end{tabular*}\label{tab1}
\end{table*}

\section{Data and Models} \label{sec:2}
\subsection{Data}\label{sec:2.1}
The observations of FRB 20240114A by FAST were carried out from 28 January to 29 August 2024, accumulating a total exposure of 33.86 hours over a 214-day campaign. The source exhibited continuous activity throughout the observation campaign, demonstrating a mean rate of 249 $\rm hr^{-1}$. A total of 11553 individual bursts were detected above the 0.026 Jy ms ($\sim12\sigma$) fluence detection threshold of FAST \citep{2025arXiv250714707Z,2025arXiv250714708Z,2025arXiv250714711Z}, which is the largest sample obtained from a single active FRB source to date. This large sample provides an opportunity to investigate the distributions of energy and waiting time, which can help reveal the properties of FRB 20240114A. 
The burst energy is calculated by 
\begin{equation}
  E=\left(10^{39} \mathrm{erg}\right) \frac{4 \pi}{1+z}\left(\frac{D_{\mathrm{L}}}{10^{28} \mathrm{~cm}}\right)^{2}\left(\frac{F_{v}}{\mathrm{Jy} \cdot \mathrm{~ms}}\right)\left(\frac{\Delta v}{\mathrm{GHz}}\right)\;,
\end{equation}
where $D_{\rm L}$ is the luminosity distance, $F_{\nu}$ is the average fluence within the frequency range specific to each burst, and $\Delta \nu$ is the bandwidth of each burst. The $D_{\rm L}$ of FRB 20240114A is estimated to be 633.87 Mpc. For energy analysis, we utilize the full data. The waiting time is defined as $\Delta t = t_{i+1} - t_i$, where $t_i$ is the arrival time of the $i$-$\rm th$ burst. Since the observation of FAST is discontinuous, with $0.3 \sim 4$ hours in each day, we pick waiting times smaller than 4 hours to discard the long observation gaps between different observation sessions. In addition, waiting times below the second valley ($\sim$ 0.5 s) of the bimodal waiting time distribution reported by \cite{2025arXiv250714707Z} are also excluded, as the right peak of the bimodal waiting time distributions of hyperactive repeaters represents the activity of the source during the statistical epoch \citep{2023ApJ...955..142Z}.

\subsection{Energy and waiting time distribution models }\label{sec2.2}
We use four models to fit the cumulative distribution function (CDF) of energy and two models to fit the CDF of waiting time. The cumulative distributions are constructed from the data as an empirical survival function (ESF), defined as $N(>X)$. Specifically, for a variable $X$ (e.g., energy $E$ or waiting time $\Delta t$) with $n$ valid data points, the sample is first sorted in ascending order as $X_1 \leq X_2 \leq \cdots \leq X_n$.
The observed cumulative number of events exceeding $X_i$ is then computed as $N_{{\rm ob},i} = N(>X_i) = n - i+1$. These models have been widely favored in previous works (e.g., \citealt{2017JCAP...03..023W,2018MNRAS.475.5109O,2018ApJ...852..140W,2019ApJ...882..108W,2020A&A...635A..61O,2020MNRAS.491.2156L,2021MNRAS.500..448C,2021ApJ...920L..23Z,2023RAA....23b5002W,2023MNRAS.519..666J,2023MNRAS.523.5430S,2024MNRAS.534.3331K,2024ApJ...974..296F}). The first model used to fit the CDF of energy is the simple power-law (SPL) distribution of the form 
\begin{equation}\label{eq 2}
 N(>E)=A\left(E^{-\alpha}-E_{{c}}^{-\alpha}\right)\;,\; E<E_{{c}}\;,
\end{equation}
where $E_c$ is the cutoff energy above which $N(> E_c)$ = 0, $\alpha$ is the power-law index. It is shown by \cite{2019ApJ...882..108W} that the CDF of energy for FRB 20121102A observed by different telescopes at different frequencies can be well fitted by the SPL distribution. The second model used to fit the CDF of energy is the bent power-law (BPL) distribution, which displays a flatter tail than the SPL distribution at low $E$, and has been applied to fit the energy CDF of FRBs \citep{2020MNRAS.491.2156L,2023RAA....23b5002W,2023MNRAS.523.5430S}.
The BPL model is given by
\begin{equation}
 N(>E)=B\left[1+(\frac{E}{E_b})^{\beta}\right]^{-1}\; , 
\end{equation}
where $B$ is the total number of bursts, $E_b$ is the median value of $E$, and $\beta$ is the power-law index. 
The third model used to fit the energy CDF is the thresholded power-law (TPL) distribution, which was originally derived by \citet{2015ApJ...814...19A} to fit the distributions of solar and stellar flares. Subsequently, \cite{2023MNRAS.523.5430S} applied the TPL model to fit the energy CDF of FRBs.
The CDF of the TPL model is given by
\begin{equation}
\begin{aligned}
N(>E) &= \int_{E}^{E_{2}} N(E)\, \mathrm{d}E \\
&= \frac{n_{0}}{1-\gamma}\bigl[\left(E_{0}+E_{2}\right)^{1-\gamma}
-\left(E_{0}+E\right)^{1-\gamma}\bigr]\;,
\end{aligned}
\end{equation}
where $n_0$ is the normalization constant, $E_0$ is the threshold parameter, and $E_2$ is the maximum value of the data. Here, $E_0$, $n_0$, and $\gamma$ are free model parameters.
The last model used to fit the energy CDF is the Band function, which was first proposed to describe the spectrum of gamma-ray bursts \citep{1993ApJ...413..281B}. It is defined as 
\begin{equation}
 N(>E)=\left\{\begin{array}{ll}
\hat{A} E^{\hat{\alpha}} \mathrm{e}^{\left(-E / \hat{E_{c}}\right)} & E \leq(\hat{\alpha}-\hat{\beta}) \hat{E_{c}} \\
\hat{A} E^{\hat{\beta}}\left[\frac{(\hat{\alpha}-\hat{\beta}) \hat{E_{c}}}{\mathrm{e}}\right]^{\hat{\alpha}-\hat{\beta}} & E \geq(\hat{\alpha}-\hat{\beta}) \hat{E_{c}}
\end{array}\right.\;,
\end{equation}
where $\hat{\alpha}$ and $\hat{\beta}$ are the power-law indices of the lower and higher energy parts of the distribution, respectively, $\hat{E_c}$ is a characteristic energy and $\hat{A}$ is the normalization factor. \cite{2022RAA....22l4002Z} showed that the Band function can provide a good fit to the CDF of FRB 20201124A. 

If FRB bursts occur randomly and independently in time and follow a Poisson process with a constant rate, then the waiting times between bursts should follow an exponential distribution. 
It has been found that the waiting times of FRBs can be described by a Poisson process \citep{2018ApJ...852..140W,2021MNRAS.500..448C,2023MNRAS.519..666J,2024ApJ...974..296F}. Therefore, the first model used to fit the CDF of waiting time is
\begin{equation}
 N(>\Delta t)\propto e^{-\lambda \Delta t}\;,
\end{equation}
where $\Delta t$ is the waiting time, $\lambda$ is the constant rate of bursts.
However, many works suggested that the repeating behavior of FRB tends to follow a Weibull distribution \citep{2018MNRAS.475.5109O,2020A&A...635A..61O,2021MNRAS.500..448C,2021ApJ...920L..23Z,2024MNRAS.534.3331K}. 
The second model used to fit the CDF of waiting time is given by 
\begin{equation}
 N(>\Delta t)\propto \mathrm{e}^{-(\Delta t r \Gamma(1+1 / k))^{k} } \;,
\end{equation}
where $\Delta t$ is the waiting time, $k$ is the shape parameter,
$r$ is the mean burst rate and $\Gamma$ is the gamma function. Note that for $k = 1$, the Weibull distribution reduces to the Poissonian case.

All the above model best-fitting parameters can be derived by minimizing 
\begin{equation}
 \chi^{2}=\sum_{i}^n \frac{\left[N_{{\rm ob},i}-N_{{\rm th},i}\right]^{2}}{\sigma_{{\rm ob},i}^{2}}\;,
\end{equation}
where $N_{{\rm ob},i}$ is the number of observed bursts, $N_{{\rm th},i}$ is the model-predicted number of bursts, $\sigma_{{\rm ob},i}$ is the uncertainty of the cumulative distribution and is taken as $\sigma_{{\rm ob},i} = \sqrt{N_{{\rm {ob}},i}} $ \citep{2019ApJ...882..108W}.

\section{Results}\label{sec3} 
The best-fitting curves of four models to the CDF of energy for FRB 20240114A are shown in the upper-left panel of Figure \ref{Fig1}, and the best-fitting parameters are listed in Table \ref{tab1}.
None of the four models can fit the full data well.
In particular, the BPL, TPL, and Band models predict values that are significantly higher than the data above $3\times10^{38}$ erg, while the SPL model predicts an energy cutoff at $\sim 5\times10^{38}$ erg. 
Furthermore, the SPL and Band models deviate from the data for energies below $4\times10^{36}$ erg. The best-fitting parameters with 1$\sigma$ uncertainties are
$\alpha = 0.276 \pm 0.001$, 
$E_{c} = (4.990 \pm 0.025) \times 10^{38}\,{\rm erg}$ 
($\chi^2_{\rm red} = 27.722$) for the SPL model; 
$\beta = 1.011 \pm 0.001$, 
$E_{b} = (1.720 \pm 0.005) \times 10^{37}\,{\rm erg}$ 
($\chi^2_{\rm red} = 5.771$) for the BPL model; 
$\gamma = 2.012 \pm 0.002$, 
$E_0 = (1.740 \pm 0.007) \times 10^{37}\,{\rm erg}$ 
($\chi^2_{\rm red} = 4.916$) for the TPL model; and 
$\hat{\alpha} = -0.258 \pm 0.001$, 
$\hat{\beta} =- 0.967 \pm 0.003$, 
$\hat{E_{c}} = (9.611 \pm 0.059) \times 10^{37}\,{\rm erg}$ 
($\chi^2_{\rm red} = 11.558$) for the Band model.
Compared with the other three models, the TPL model provides the best fit to the CDF of energy because it has the smallest reduced chi-square value. We also show the fit of the SPL model after excluding the data below the 90\% detection completeness threshold of FAST, which yields the best-fitting parameters with 1$\sigma$ uncertainties $\alpha = 0.373 \pm 0.001$, 
$E_{c} = (6.389 \pm 0.026) \times 10^{38}\,{\rm erg}$ 
($\chi^2_{\rm red} = 10.523$).

The best-fitting curves of the Poisson and Weibull models to the CDF of waiting time for FRB 20240114A are shown in the upper-right panel of Figure \ref{Fig1}, and the best-fitting parameters are listed in Table~\ref{tab1}. The best-fitting parameter for the Poisson model is $\lambda = (0.919\pm0.001)\times10^{-1}\rm s^{-1}$ (1$\sigma$ uncertainty), and the best-fitting parameters for the Weibull model are $k =0.777\pm0.001$ and $r=(0.989\pm0.001)\times10^{-1}\rm s^{-1}$ (1$\sigma$ uncertainty). As we can see, both the Poisson and Weibull models predict values that are significantly lower than the data for waiting times greater than 40 s. The Weibull model fits the CDF of waiting time better than the Poisson model because the reduced chi-square value of the Weibull model ($\chi^2_{\rm red}=3.529$)
is smaller than that of the Poisson model ($\chi^2_{\rm red}=11.993$). Nonetheless, neither model provides a good fit to the waiting time distribution.

We note that fitting cumulative distributions $N(>X)$ introduces correlated errors, which reduce the effective number of degrees of freedom and affect the interpretation of the reduced $\chi^2$. To assess the robustness of our results, we also perform an independent, unbinned maximum-likelihood analysis of the energy and waiting time distributions (see details in Appendix \ref{app}).
The results shown in the lower panels of Figure \ref{Fig1} indicate that none of the models provides a satisfactory fit to the full dataset.

In summary, for the full energy and waiting time sample of FRB 20240114A, a single model cannot fit the data well. The full energy sample was also fitted with log‑normal and bimodal log‑normal (Bi‑LN) models by \cite{2025arXiv250714707Z}, who reported that the Bi‑LN model provides a better description. This may indicate the presence of multiple types of bursts in FRB 20240114A. 
Therefore, we treat the bursts from each single-day observation session as a subsample and analyze them separately. 

\begin{figure*}
\centering
\includegraphics[width=0.45\textwidth]{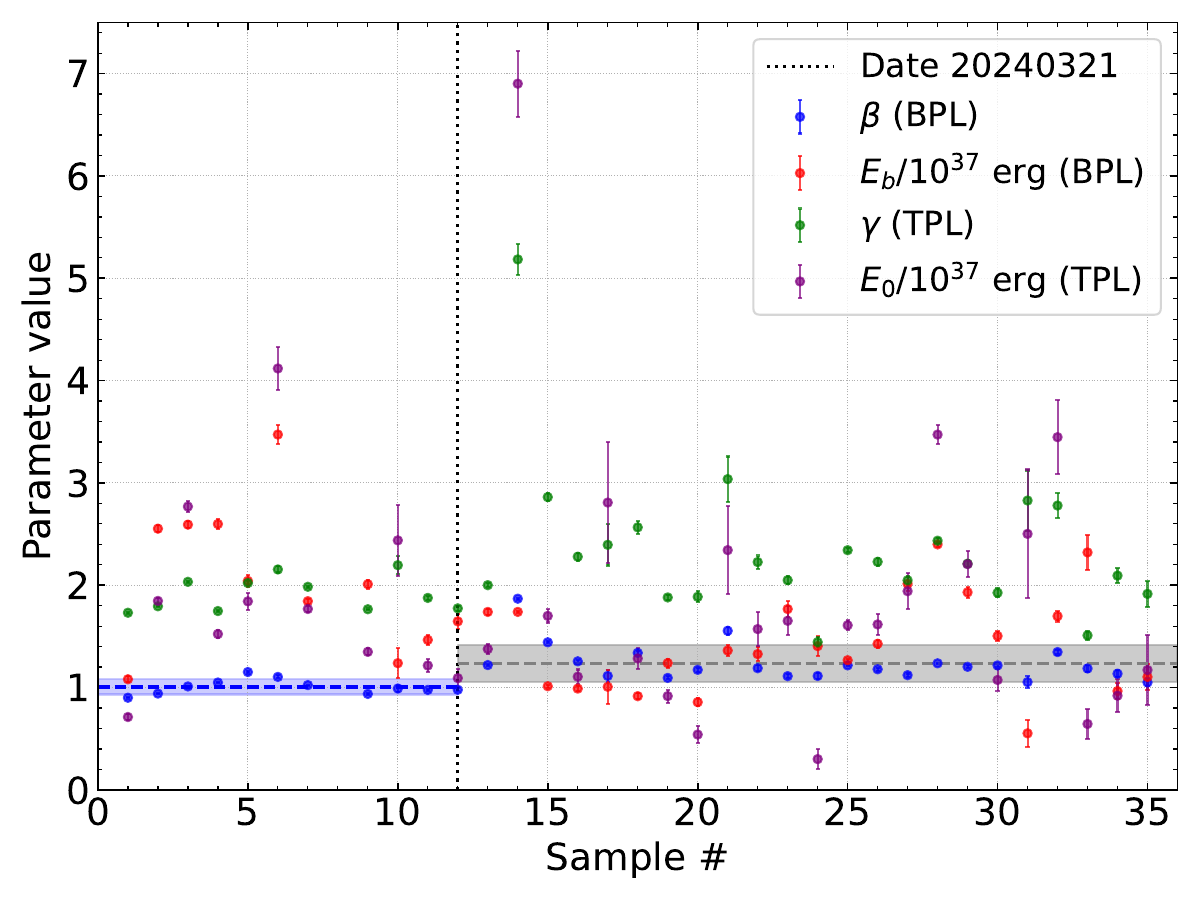}
\hspace{0.03\textwidth} 
\includegraphics[width=0.45\textwidth]{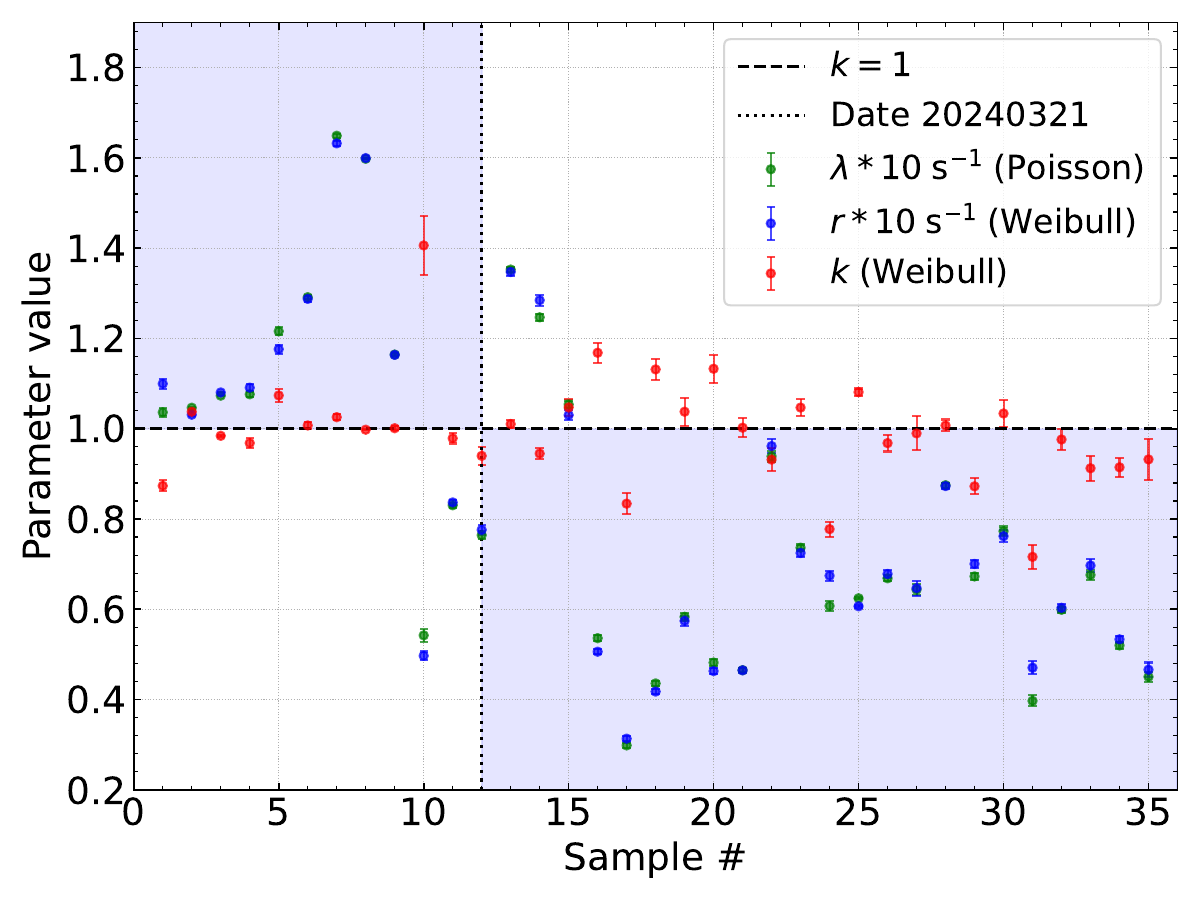}
\caption{The best-fitting parameters for FRB 20240114A in single-day observation sessions. The left panel shows the BPL model parameters $\beta$ and $E_b$ and TPL model parameters $\gamma$ and $E_0$ for the energy distribution. The dashed blue line with the blue shaded region and the dashed gray line with the gray shaded region display the mean and standard deviation of the best-fitting BPL parameter $\beta$ for the samples observed before and after 21 March 2024, respectively. 
The right panel shows the Weibull parameters $k$ and $r$ and Poisson parameter $\lambda$ for the waiting time distribution. The Weibull parameter $r$ and the Poisson parameter $\lambda$ for most subsamples observed after 21 March 2024 are lower than those for subsamples observed before 21 March 2024.}\label{Fig4}
\end{figure*}

\subsection{Energy distribution in single-day observation sessions}\label{sec3.1}

Over a roughly seven-month period, FAST carried out 57 observations of FRB 20240114A. We select single-day observations with a number of detected bursts larger than 50 to investigate the energy distribution. 
A total of 35 days with more than 50 bursts are identified, thus resulting in 35 subsamples.
We mark each subsample by its observation date, e.g., ``Date 20240312'' denotes the bursts observed on 12 March 2024. The number of bursts in each subsample is shown at the top of each panel in Figure \ref{Fig2}. We use the SPL, BPL, and TPL models to fit the CDF of energy for each subsample. The Band function includes one more parameter than the other three models, so we do not adopt this more complex model. 
The best-fitting curves of the CDFs of energy for the 35 subsamples are shown in Figure \ref{Fig2}.
The BPL and TPL models provide a better description of the energy distribution for most subsamples compared to the SPL model.
However, for the subsample Date 20240312, none of the three models provides a good fit. Additionally, we note that most bursts with energies above $10^{39}$ erg occurred before 21 March 2024. The best-fitting parameters for the BPL and TPL models are summarized in Table \ref{tab2}. The parameter values $\beta$, $E_b$, $\gamma$ and $E_0$ as functions of subsample order are shown in the left panel of Figure \ref{Fig4}. 
One can see that the best-fitting TPL parameter $\gamma$ ranges from 1.5 to 5.5, $E_0$ is distributed in the range of $ (0.3-7)\times10^{37}$ erg. The best-fitting BPL parameter $\beta$ ranges from $0.9$ to $2.0$, $E_b$ is distributed in the range of $ (0.5-3.5)\times10^{37}$ erg. Interestingly, the best-fitting parameter $\beta$ of the BPL model for the subsamples observed after 21 March 2024 appears to be slightly larger than for the subsamples observed before 21 March 2024, and $\beta$ is approximately invariant within each epoch. The mean and standard deviation are $\bar \beta_b = 1.006 \pm 0.074$ and $\bar \beta_a = 1.236 \pm 0.183$, respectively.
The BPL is a smoothly connected piecewise power law that has a flat tail at the low-energy end ($E<E_b$), and reduces to a simple power law at the high-energy end ($E>E_b$).
This implies that the energy statistics of FRB 20240114A show different properties before and after 21 March 2024, which may be caused by two distinct burst types.

\subsection{Waiting time distribution in single-day observation sessions}\label{sec3.2}
We select single-day observations with more than 50 detected bursts as subsamples, and exclude the data with waiting times shorter than 0.5 s for each subsample to analyze the waiting time distribution. Similar to the energy subsamples, the 35 waiting time subsamples were identified according to the observing date. The number of waiting times in each subsample is shown at the top of each panel in Figure \ref{Fig3}. The best-fitting curves of the CDFs of waiting time for the 35 subsamples are shown in Figure \ref{Fig3}, and the best-fitting parameters for the Poisson and Weibull models are summarized in Table \ref{tab2}.
The Poisson and Weibull models both describe the CDF of waiting time for all subsamples well. 
However, compared to the Poisson model, the Weibull model with a $k$ parameter that deviates from 1 provides a better fit to the data. The right panel of Figure \ref{Fig4} shows the parameters $k$, $r$, and $\lambda$ as a function of the subsample order. 
It can be seen that the best-fitting Weibull parameter $k$ ranges from $0.7$ to $1.4$, and both the best-fitting Weibull parameter $r$ and Poisson parameter $\lambda$ are distributed in the range of $(0.03-0.18) \;\rm{s^{-1}}$. 
Although the best-fitting results vary for different subsamples, the Poisson parameter $\lambda$ is nearly consistent with the Weibull parameter $r$ for most subsamples. 
The best-fitting parameter $k$ of the Weibull model is close to 1 for the subsamples Date 20240310, 20240312, 20240313, 20240327, 20240512 and 20240729.
For $k = 1$, the Weibull distribution reduces to the Poissonian case, indicating that the burst activity of FRB 20240114A is random on these days. In addition, we note that the Weibull parameter $r$ and the Poisson parameter $\lambda$ for most subsamples observed after 21 March 2024 are lower than those for subsamples observed before 21 March 2024. This implies that the burst activity in the earlier epoch is more intense than in the later epoch.

\section{Discussions }\label{sec4}
\begin{figure}
\centering
\includegraphics[width=0.45\textwidth]{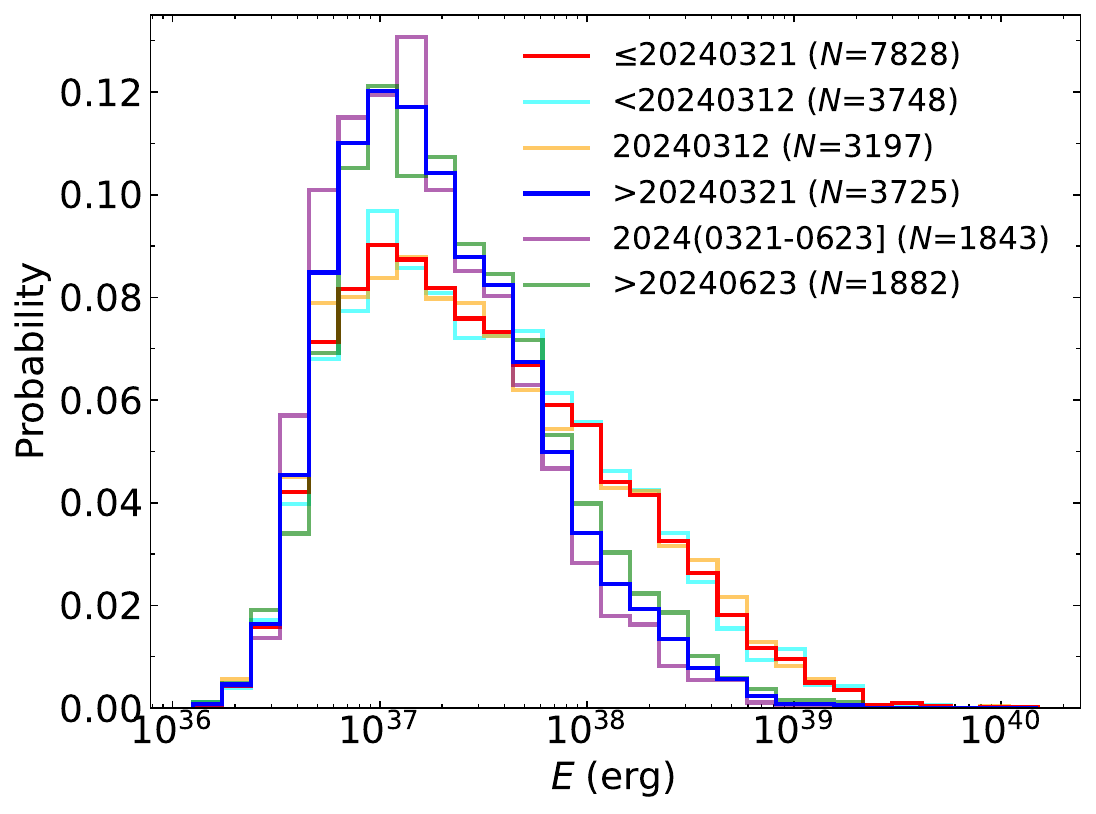}
\caption{The probability histograms of energy 
for the six large subsamples. The number of bursts for each subsample is indicated in the upper-right corner of the figure panel.
}\label{Fig5}
\end{figure}

In Section \ref{sec3.1}, we find that most high-energy bursts with energies above 
$10^{39}$ erg were detected before 21 March 2024, and the CDFs of energy for the subsamples detected before and after this date show significant differences. Therefore, we further examine the statistical distributions of energy in different observation epochs. The six large subsamples are generated based on different observation epochs for comparison. First, we divide the full data into two subsamples: bursts observed on and before 21 March 2024 (marked $\leq$20240321) and bursts observed after 21 March 2024 (marked $>$20240321). Second, since the number of bursts detected on 12 March 2024 was markedly higher than on any other single-day observation, we treat the bursts observed on that day as an independent subsample (marked 20240312). Third, to avoid the statistical weighting bias introduced by the 12 March 2024 observation, we define all bursts observed before 12 March 2024 (marked $< $20240312) as another subsample. Finally, the bursts observed after 21 March 2024 are divided into two subsamples with nearly equal numbers of bursts, denoted $2024(0321$-$0623]$ and $>$20240623, respectively. The six large energy subsamples of $\leq$20240321, $<$20240312, 20240312, $>$20240321, $2024(0321$-$0623]$ and $>$20240623 contain 7828, 3748, 3197, 3725, 1843, and 1882 bursts, respectively. 
Figure \ref{Fig5} shows the probability distributions of energy for six large subsamples.
The $p$-values and D-statistics from the Kolmogorov–Smirnov (KS) tests for the six large subsamples are listed in Table \ref{p_value2}.
It is clear that the subsamples observed before and after 21 March 2024 show significant differences in their energy probability distributions, particularly at the high-energy end ($>6\times10^{37}$ erg). Although the two subsamples observed after 21 March 2024 ($2024(0321$-$0623]$ and $>$20240623) still exhibit statistically significant differences ($p = 5.82\times10^{-7}$), the two subsamples observed before this date ($<$20240312 and 20240312) are statistically consistent ($p = 0.21$).

\begin{figure}
\centering
\includegraphics[width=0.45\textwidth]{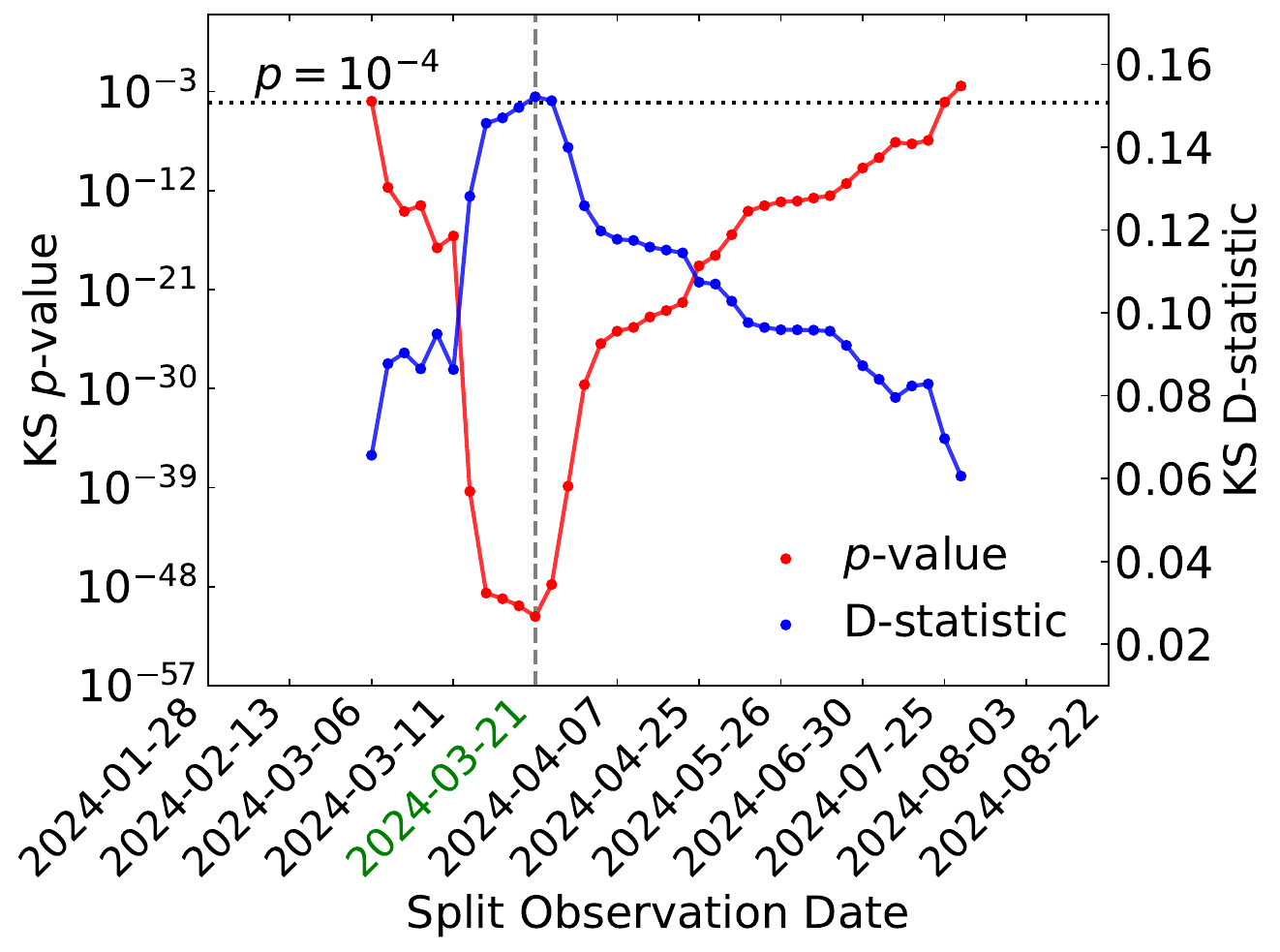}
\includegraphics[width=0.45\textwidth]{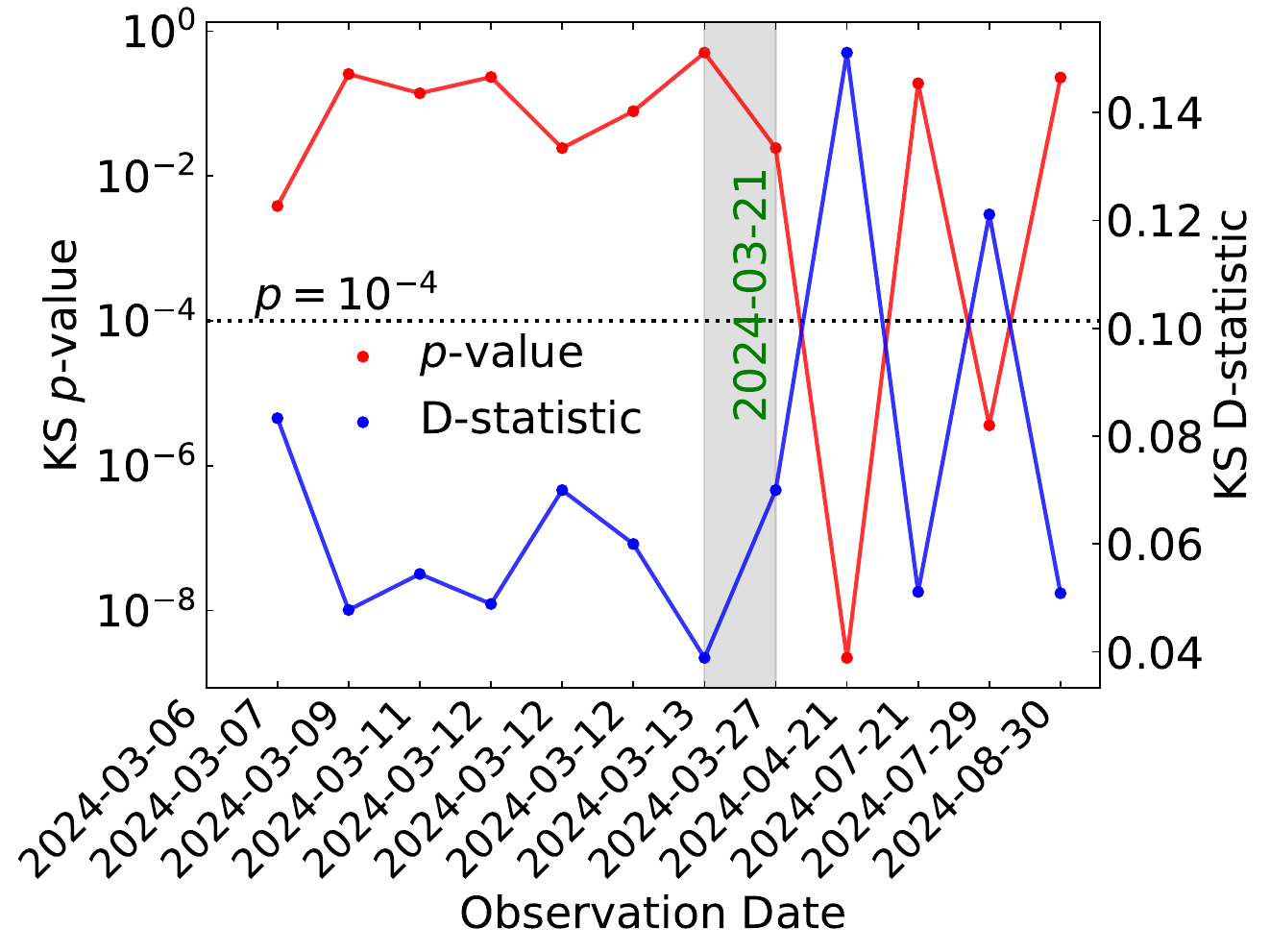}
\caption{
{\em Upper panel---} The $p$-values ({\em red circles}) and D-statistics ({\em blue circles}) of the KS test between two energy samples obtained by splitting the full dataset at each observation date as a function of split observation date.
{\em Lower panel---} The $p$-values ({\em red circles}) and D-statistics ({\em blue circles}) of the KS tests between adjacent time-ordered subsamples of bursts as a function of observation date.}\label{Fig6}
\end{figure}

We perform a more detailed investigation of the dependence of the differences in the energy distributions on the split observation date. The full energy data are split by observation date into two samples, with each sample containing at least 900 bursts.
The upper panel of Figure \ref{Fig6} shows the $p$-values and D-statistics of the KS test between the two energy samples as functions of the split observation date. The results clearly show that the most significant differences occur between the two samples observed before and after 21 March 2024. Meanwhile, we further examine the temporal evolution of the differences in the energy distributions. Specifically,
the full energy data are divided into time-ordered subsamples, each containing 900 bursts. The representative time of each subsample is taken as the observation date of the last burst within the corresponding time window. For example, the first subsample spans the time window from 28 January 2024 to 6 March 2024, and its representative time is taken as 6 March 2024. KS tests are then performed between adjacent subsamples.
The lower panel of Figure \ref{Fig6} shows the $p$-values and D-statistics of the KS tests between adjacent subsamples as a function of the observation date, where the observation date is adopted as the representative time of the latter of the two adjacent subsamples. It is evident that the subsamples observed before 21 March 2024 are relatively consistent, whereas those observed after this date exhibit more complex variability.
We also investigate whether the waiting time distributions for the two epochs also show differences. The solid and dashed red lines in Figure \ref{Fig8} show the kernel density estimate (KDE) of waiting times of bursts observed before and after 21 March 2024, with medians of 5.87 s and 11.34 s, respectively.
The bursts detected in the latter epoch have a longer median waiting time, implying lower source activity than in the earlier epoch. The observed differences in energy and waiting time distributions between the two epochs may indicate that distinct burst types dominate in each epoch, possibly due to changes in the physical properties of the emission region.

\begin{figure}
\centering
\includegraphics[width=0.45\textwidth]{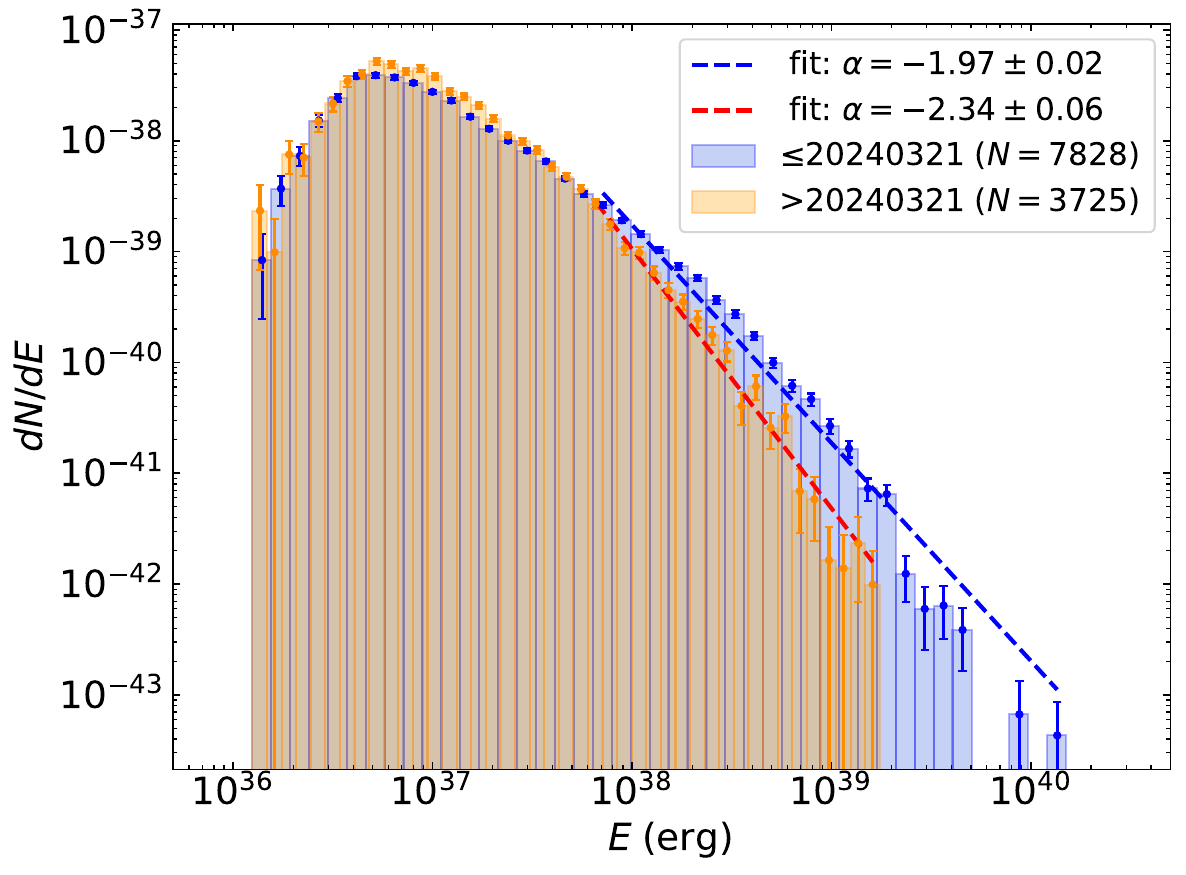}
\caption{The differential energy distribution of bursts from FRB 20240114A observed before and after 21 March 2024. The dashed blue and red lines show the best-fitting results for two epochs, respectively.}\label{Fig7}
\end{figure}

\begin{figure}
\centering
\includegraphics[width=0.45\textwidth]{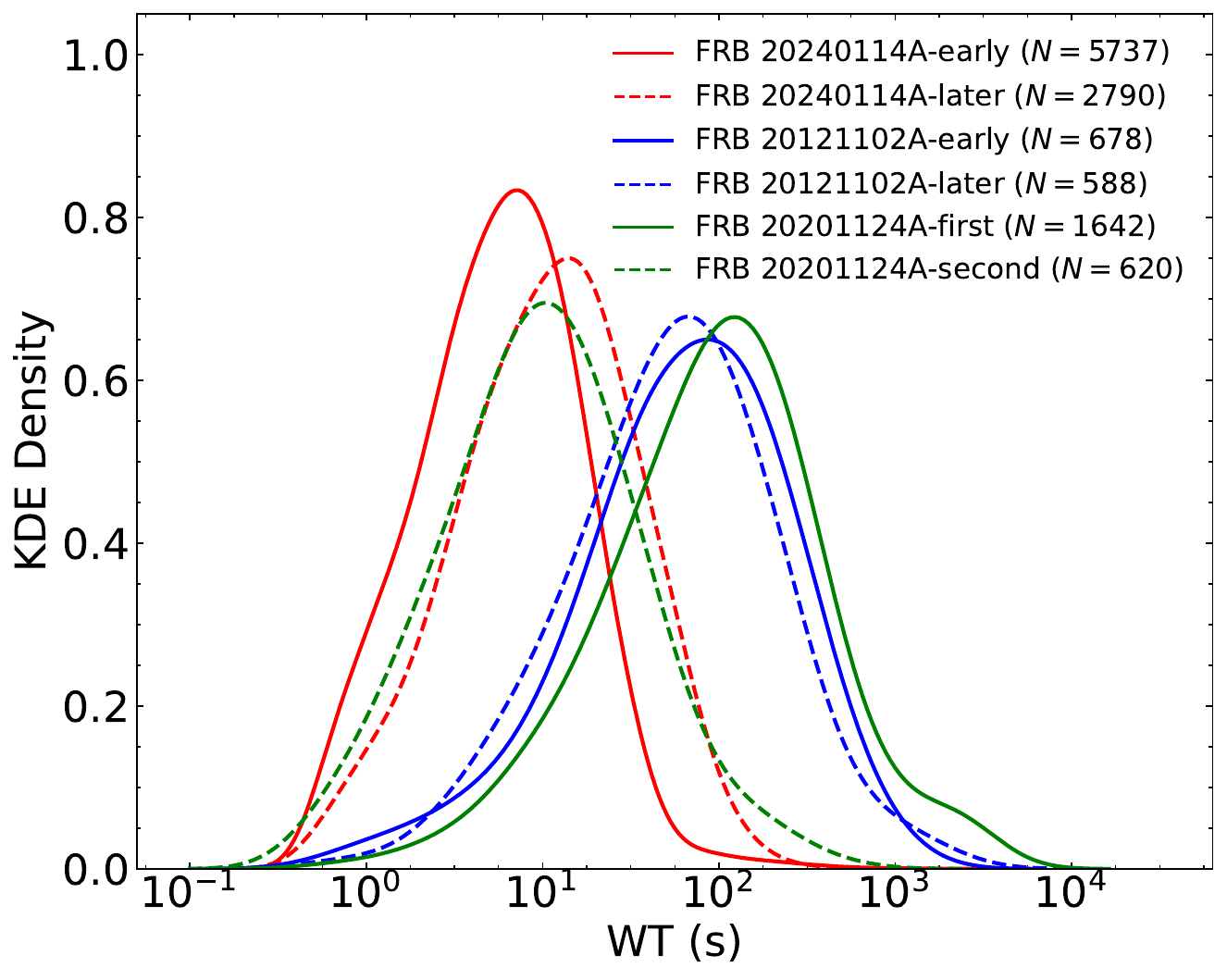}
\caption{
The KDE of waiting times for FRB 20121102A, FRB 20201124A and FRB 20240114A in different observation epochs. The number of waiting times for each subsample is shown in the upper-right corner of the figure panel.}\label{Fig8}
\end{figure}

The variations in the energy distribution of FRB 20240114A in different epochs show certain similarities with other hyperactive repeaters, whereas its waiting time distributions exhibit significant differences. 
Similar to the significant differences in energy distributions observed in FRB 20121102A before and after MJD 58740 \citep{2021Natur.598..267L}, power-law fits to the differential energy distribution $dN/dE$ for bursts with energies $E>10^{38}$ erg yield indices of $-1.70_{-0.03}^{+0.03}$ and $-2.60_{-0.14}^{+0.15}$ (1$\sigma$ uncertainties), respectively \citep{2021ApJ...920L..23Z}. The differential energy distributions of FRB 20240114A bursts observed before and after 21 March 2024 are shown in Figure \ref{Fig7}. Power-law fits to bursts with $E > 6\times10^{37}$ erg yield indices of $-1.97_{-0.02}^{+0.02}$ and $-2.34_{-0.06}^{+0.06}$, respectively. In contrast, the waiting time evolution in FRB 20240114A is distinct from that observed in FRB 20121102A and FRB 20201124A. Figure \ref{Fig8} also presents the KDE of the burst waiting times for the early and late epochs of FRB 20121102A, separated at MJD 58740 following \cite{2021ApJ...920L..23Z} (solid and dashed blue lines), and for the two active episodes of FRB 20201124A reported by \cite{2022Natur.609..685X} and \cite{2022RAA....22l4002Z} (solid and dashed green lines), where the waiting times with WT $< 30$ ms and WT $ > 1$ hour are ignored. For these sources, bursts in the later active epochs generally have shorter median waiting times. However, FRB 20240114A shows the opposite behavior, with the median waiting time increasing from 5.87 s in the early epoch to 11.34 s in the late epoch. These results suggest that bursts from different hyperactive repeaters may be driven by different physical processes.

\begin{table}
\centering
\caption{
The $p$-values and D-statistics of the KS tests  between Large Subsamples.}
\begin{tabular}{cccc}
\hline\hline
subsample 1 &subsample 2 & \textit{p}  &  D     \\
\hline
$\leq$20240321 & $>$20240321        & $<10^{-8}$  &  0.152 \\
$\leq$20240321 & 2024(0321-0623$]$  & $<10^{-8}$  &  0.185   \\
$\leq$20240321 & $>$20240623        & $<10^{-8}$  &  0.126    \\
$>$20240321    & $<$20240312        & $<10^{-8}$  &  0.159    \\
$>$20240321    & 20240312           & $<10^{-8}$  &  0.153    \\
$<$20240312    & 20240312           & $0.21$   &  0.026     \\
$<$20240312    & 2024(0321-0623$]$  & $<10^{-8}$  &  0.195      \\
$<$20240312     & $>$20240623        & $<10^{-8}$  &  0.129     \\
20240312       & 2024(0321-0623$]$  & $<10^{-8}$  &  0.183        \\
20240312       & $>$20240623        & $<10^{-8}$  &  0.127       \\
$>$20240623    & 2024(0321-0623$]$  & $5.82\times10^{-7}$   &  0.090         \\
\hline
\end{tabular}\label{p_value2}
\end{table}

\section{Conclusions }\label{sec5}
In this paper, we present a comprehensive analysis of energy and waiting time distributions of FRB 20240114A observed by FAST. We use the SPL, BPL, TPL, and Band models to fit the CDF of energy, and apply the Poisson and Weibull models to fit the CDF of waiting time (with WT $>$ 0.5 s and WT $<$ 4 hours) for the full sample. Our results indicate that no single model can fit the distribution of energy or waiting time well. We select 35 subsamples based on single-day observations, each containing more than 50 bursts, and fit their energy distributions with SPL, BPL, and TPL models and their waiting time distributions with the Poisson and Weibull models. The results show that the BPL and TPL models describe the CDFs of energy for most subsamples well, except for the subsample observed on 12 March 2024. The best-fitting parameter $\beta$ of the BPL model is approximately invariant within each epoch (before and after 21 March 2024). The mean and standard deviation are $\bar \beta_b = 1.006 \pm 0.074$ and $\bar \beta_a = 1.236 \pm 0.183$, respectively. The CDFs of waiting time for all subsamples are better described by a Weibull model. Although the best-fitting results vary for different subsamples, the Weibull parameter $r$ for most subsamples observed after 21 March 2024 is lower than for those observed before that date. Almost all high-energy bursts with energies above $10^{39}$ erg occurred in observations before 21 March 2024. The energy probability distributions of the burst samples observed before and after 21 March 2024 show the most significant differences, particularly at the high-energy end ($E > 6\times10^{37}$ erg). We use the power-law to fit the $dN/dE$ of bursts with energy greater than $ 6\times10^{37}$ erg. The fitted power law indices are $-1.97_{ - 0.02}^{ + 0.02}$ and $-2.34_{ - 0.06}^{ + 0.06}$, respectively. The waiting time distributions for the two epochs also differ markedly, with medians of 5.87 s and 11.34 s for the earlier and later epochs, respectively. These results indicate that the two epochs may be characterized by different types of bursts, possibly attributed to changes in the physical properties of the emission region.

\section*{acknowledgments} 
We thank the referee for helpful and insightful comments that improved the quality of the manuscript.
This work is supported by the National Key R\&D Program (No. 2024YFA161100) , the Guangxi Science and Technology Innovation Platform Program (Leitai Action Plan, Grant No. Guike LT2600640026 ), Guangxi Key R\&D Program (Guangxi Funeng Action Plan, Grant No. FN2504240030), National Natural Science Foundation of China (grant Nos. 12133003),  and the Innovation Project of Guangxi Graduate Education (grant No. YCBZ2026029). E. W. L. is also supported by the Guangxi Talent Program (``Highland of Innovation Talents'').

\section*{Data Availability}
The data of FRB 20240114A are available in \citealt{2025arXiv250714707Z}.

\bibliographystyle{mnras}
\bibliography{example}

\appendix
\section{Unbinned Maximum-Likelihood Analysis of the energy and waiting time distributions}\label{app}

To assess the robustness of our results obtained from fitting cumulative distributions, we perform an independent, unbinned maximum-likelihood analysis of the energy and waiting time distributions. This approach avoids the binning effects inherent in fitting binned differential distributions and also mitigates the correlated errors that arise in cumulative distributions.

Similar to Section \ref{sec2.2}, we adopt four models to fit the energy distribution.
The first model is
\begin{equation}
\frac{dN}{dE}= E^{-(\alpha+1)} \exp(-E/E_{c})\;,
\end{equation}
where $\alpha$ is the power-law index and $E_{c}$ is the cutoff energy. Note that this form differs from the direct derivative of Equation \ref{eq 2}. We include the exponential cutoff term in the model to provide a smooth suppression at the high‑energy end and to improve the stability of the fit.
The second model is given by  \begin{equation}
\frac{dN}{dE}= 
\frac{\beta \left(E/E_{b}\right)^{\beta-1}}{E_{b}\left[1+\left(E/E_{b}\right)^{\beta}\right]^2}\;,
\end{equation}
where $\beta$ is the power-law index and $E_{b}$ is the median value of $E$.
The third model is given by 
\begin{equation}
\frac{dN}{dE} = n_0(E + E_0)^{-\gamma}\;,
\end{equation}
where $n_0$ is the normalization constant, $\gamma$ is the power-law index, and $E_0$ is the threshold parameter.
The final model is given by 
\begin{equation}
\frac{dN}{dE} =\left\{\begin{array}{ll}
 E^{\hat{\alpha}} \mathrm{e}^{\left(-E / \hat{E_{c}}\right)} & E \leq(\hat{\alpha}-\hat{\beta}) \hat{E_{c}} \\ E^{\hat{\beta}}\left[\frac{(\hat{\alpha}-\hat{\beta}) \hat{E_{c}}}{\mathrm{e}}\right]^{\hat{\alpha}-\hat{\beta}} & E \geq(\hat{\alpha}-\hat{\beta}) \hat{E_{c}}
\end{array}\right.\;,
\end{equation}
where $\hat{\alpha}$ and $\hat{\beta}$ are the power-law indices of the lower and higher energy parts of the distribution, respectively, and $\hat{E_{c}}$ is a characteristic energy.

We use two models to fit the waiting time distribution. The first model is given by
\begin{equation}
 \frac{dN}{d(\Delta t)}= \lambda e^{-\lambda \Delta t}\;,
\end{equation}
where $\lambda$ is the constant rate of bursts.
The second model is given by
\begin{equation}
\frac{dN}{d(\Delta t)} = k \left( r\,\Gamma\!\left(1 + {1}/{k}\right) \right)^k
\Delta t^{k-1}
\exp\left[ - \left( r\,\Delta t\,\Gamma\!\left(1 +{1}/{k}\right) \right)^k  \right] \;,
\end{equation}
where $k$ is the shape parameter,
$r$ is the mean burst rate, and $\Gamma$ is the gamma function.

The best-fitting parameters for all models are obtained by minimizing the negative log-likelihood:
\begin{equation}
- \ln \mathcal{L} = - \sum_{i=1}^{N} \ln p(X_i)\;,
\end{equation}
where $p(X_i)$ denotes the normalized probability density function derived from the model $dN/dE$ or $dN/d(\Delta t)$. The normalization is computed numerically over the range $[X_{\rm min}, X_{\rm max}]$.
The goodness of fit is assessed using a Kolmogorov–Smirnov (KS) test as a diagnostic tool, by comparing the empirical cumulative distribution of the unbinned data with the best-fitting model distributions.
The best-fitting curves of the four models for the energy distribution of FRB 20240114A are shown in the lower left panel of Figure \ref{Fig1}, while those of the two models for the waiting time distribution are shown in the lower right panel. 
We find that none of the models provides a satisfactory fit to the observed data, consistent with the results obtained from the cumulative distribution analysis.

\section{Fitting Results for Burst Energies and Waiting Times in Single-Day Subsamples}
For completeness, we show the best-fitting cumulative distribution functions (CDFs) of burst energies for the 35 subsamples in Figure \ref{Fig2}, while those of waiting times are presented in Figure \ref{Fig3}.
The best-fitting parameters of the BPL and TPL models for burst energies, and the Poisson and Weibull models for waiting times, are summarized in Table \ref{tab2}.

\begin{figure*}
\centering
\includegraphics[width=\textwidth,keepaspectratio]{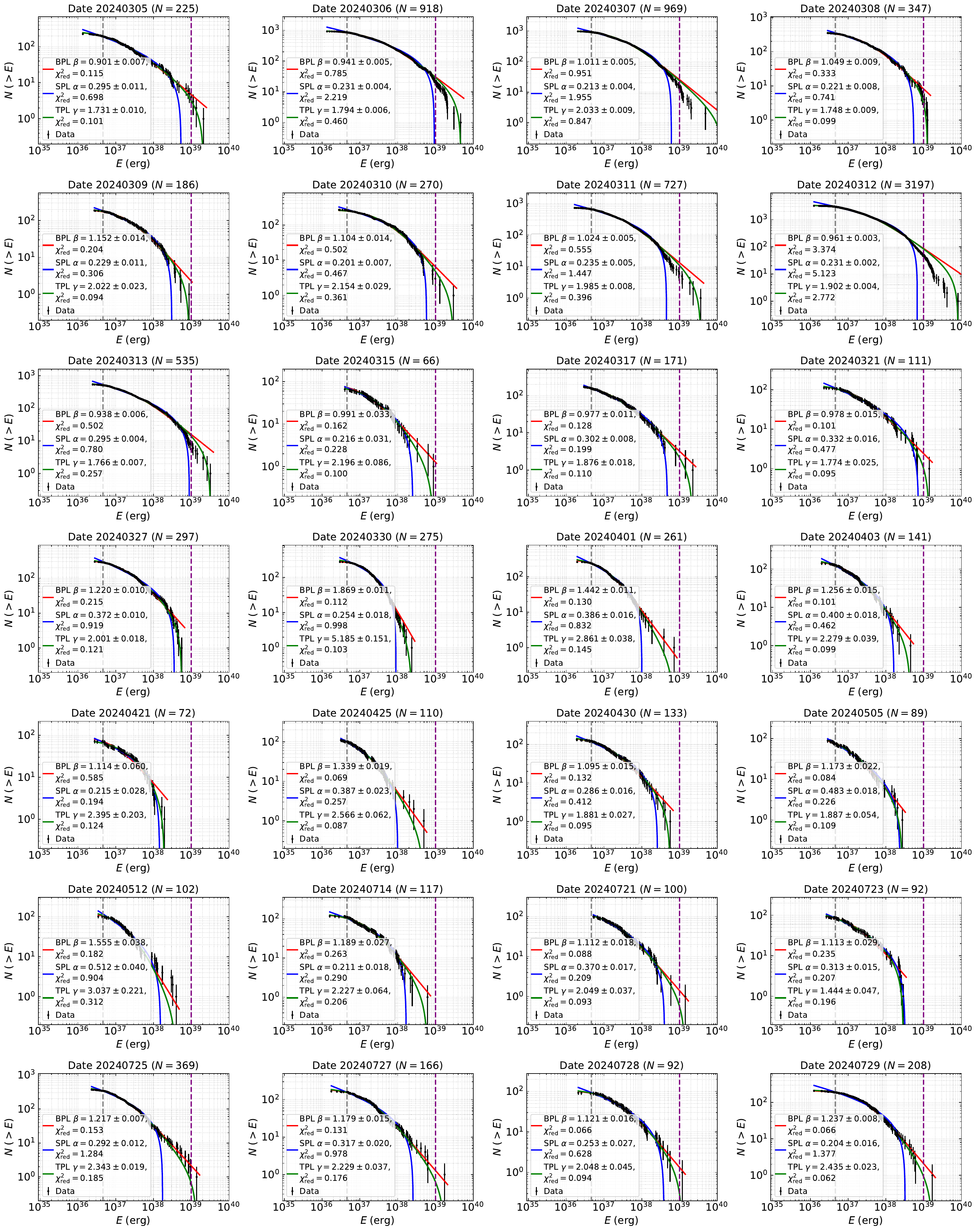}
\caption{The CDFs of energy for FRB 20240114A in single-day observation sessions, with the blue, red, and green solid lines indicating the best-fitting curves for SPL, BPL and TPL models, respectively. The gray dashed line indicates the 90\% detection completeness threshold of FAST. The dashed purple line corresponds to bursts with an energy of $10^{39}$ erg.}\label{Fig2}
\end{figure*}

\begin{figure*}
\ContinuedFloat
\centering
\includegraphics[width=\textwidth,keepaspectratio]{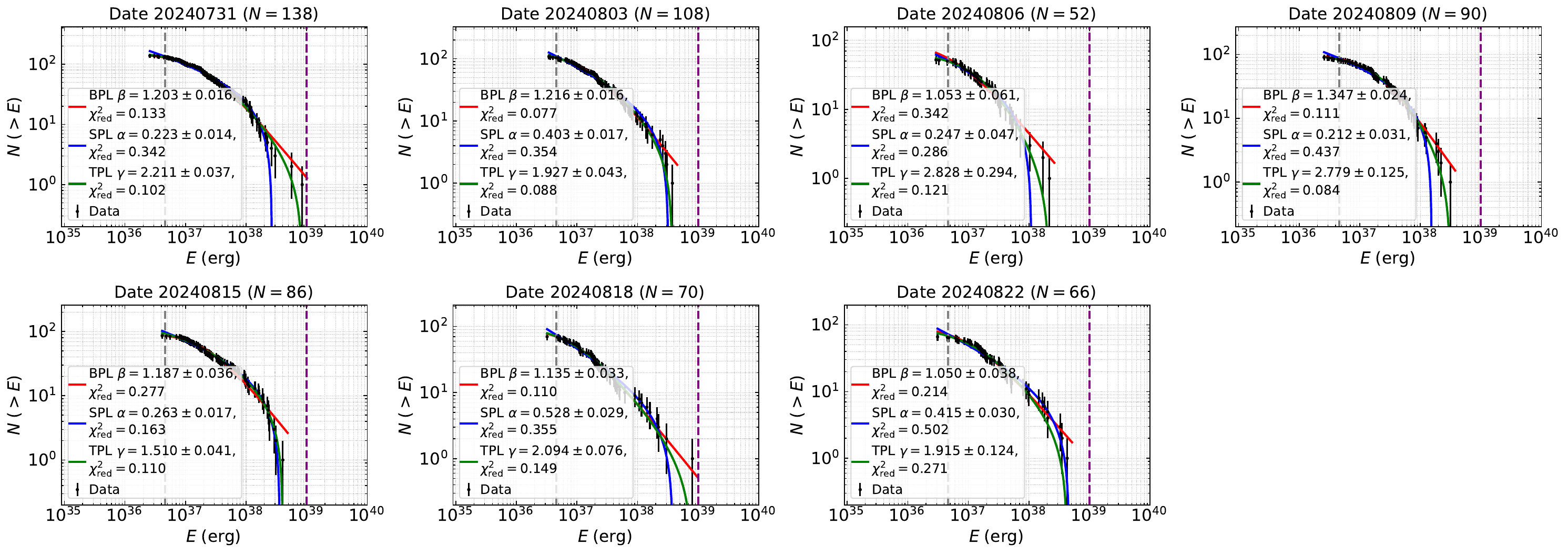}
\caption{continued}
\end{figure*}

\begin{figure*}
\centering
\includegraphics[width=\textwidth,keepaspectratio]{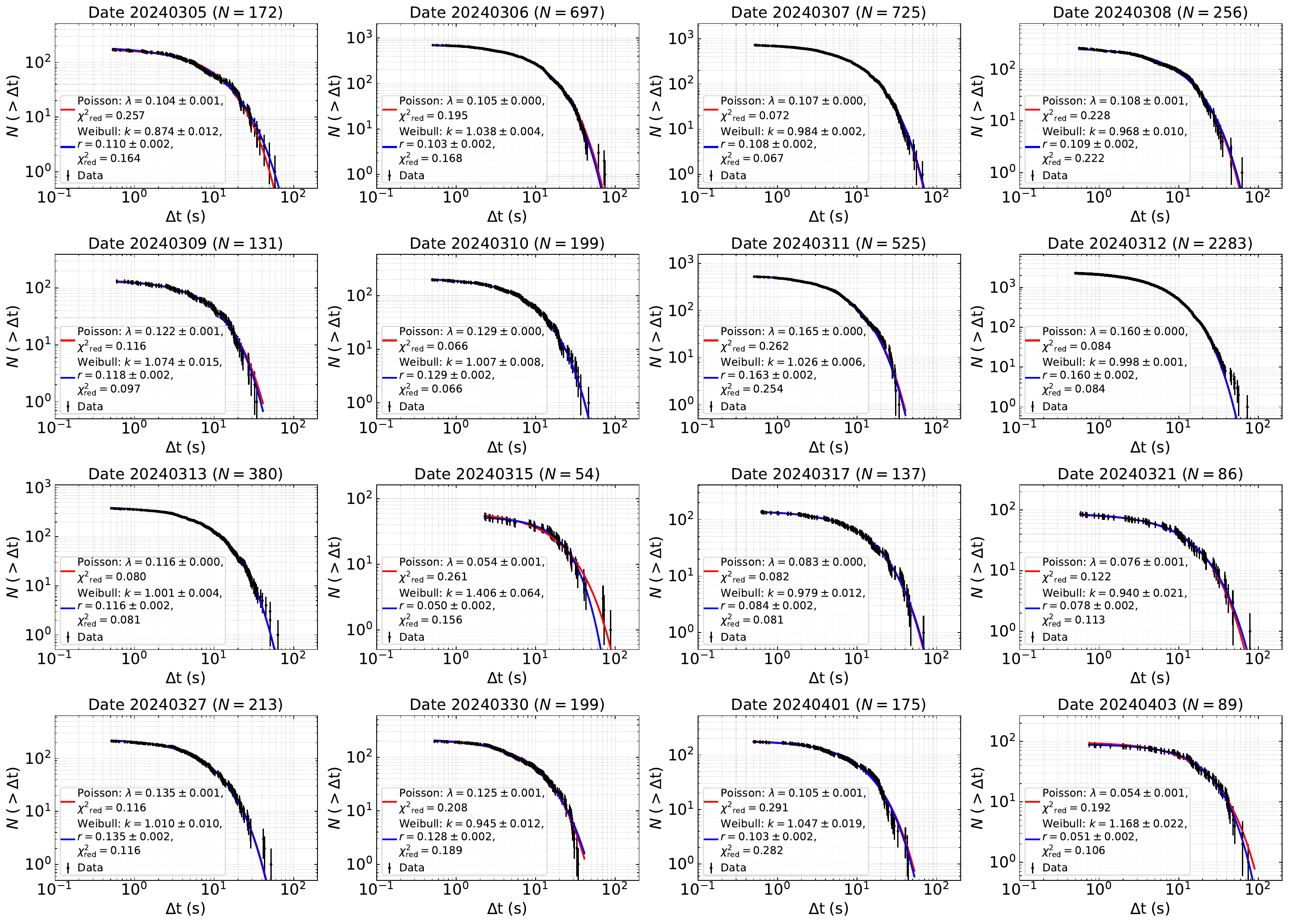}
\caption{The CDFs of waiting time for FRB 20240114A in single-day observation sessions, with the red and blue solid lines representing the best-fitting curves for the Poisson and Weibull models, respectively. }\label{Fig3}
\end{figure*}

\begin{figure*}\ContinuedFloat
\centering
\includegraphics[width=\textwidth,keepaspectratio]{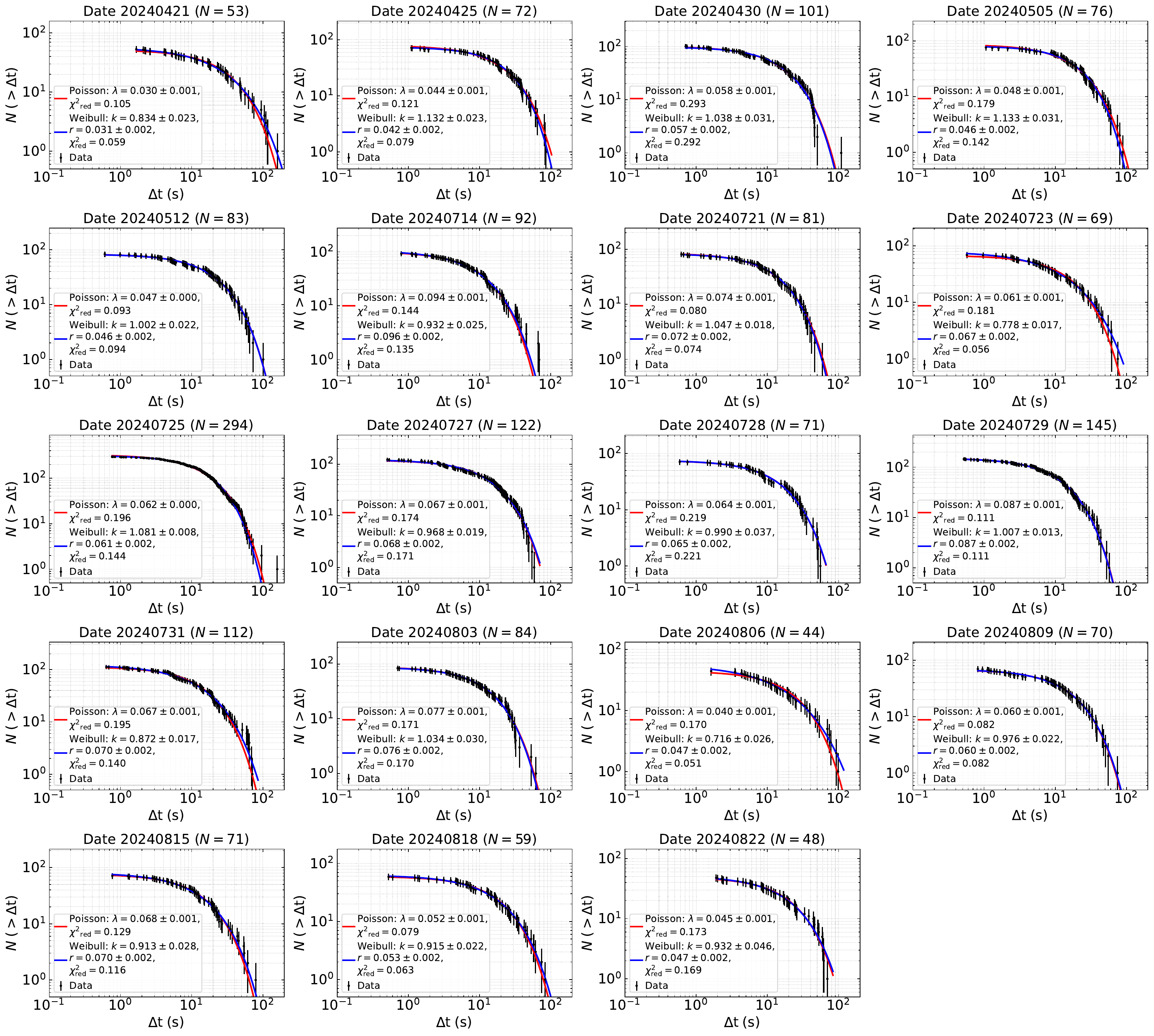}
\caption{continued}
\end{figure*}

\begin{table*}
\centering
\caption{The best-fitting parameters for the BPL and TPL models of the energy CDF, and for the Poisson and Weibull models of the waiting time CDF of FRB 20240114A in single-day observation sessions.}
\label{tab2}
\begin{adjustbox}{width=1\textwidth} 

\centering
\begin{adjustbox}{width=1\textwidth} 
\begin{tabular}{llccc ccc ccc cc}  
\toprule
 & & \multicolumn{3}{c}{BPL (energy)} & \multicolumn{3}{c}{TPL (energy)} & \multicolumn{3}{c}{Weibull (WT)} & \multicolumn{2}{c}{Poisson (WT)} \\
\cmidrule(lr){3-5}\cmidrule(lr){6-8}\cmidrule(lr){9-11}\cmidrule(lr){12-13}
No. & burst date & $\beta$ & $E_b\,[10^{37}\,\mathrm{erg}]$ & $\chi^2_{\rm red}$ 
                & $\gamma$ & $E_0\,[10^{37}\,\mathrm{erg}]$ & $\chi^2_{\rm red}$
                & $k$ & $r\,[10^{-1}\,\mathrm{s}^{-1}]$ & $\chi^2_{\rm red}$
                & $\lambda\,[10^{-1}\,\mathrm{s}^{-1}]$ & $\chi^2_{\rm red}$ \\
\midrule
1  & 20240305 & $0.901\pm0.007$ & $1.081\pm0.028$ & 0.115 & $1.731\pm0.010$ & $0.712\pm0.028$ & 0.101 & $0.874\pm0.012$ & $1.098\pm0.011$ & 0.164 & $1.036\pm0.009$ & 0.257 \\
2  & 20240306 & $0.941\pm0.005$ & $2.553\pm0.034$ & 0.785 & $1.794\pm0.006$ & $1.847\pm0.032$ & 0.460 & $1.038\pm0.004$ & $1.031\pm0.002$ & 0.168 & $1.047\pm0.002$ & 0.195 \\
3  & 20240307 & $1.011\pm0.005$ & $2.592\pm0.033$ & 0.951 & $2.033\pm0.009$ & $2.769\pm0.054$ & 0.847 & $0.984\pm0.002$ & $1.080\pm0.001$ & 0.067 & $1.073\pm0.001$ & 0.072 \\
4  & 20240308 & $1.049\pm0.009$ & $2.598\pm0.053$ & 0.333 & $1.748\pm0.009$ & $1.523\pm0.043$ & 0.099 & $0.968\pm0.010$ & $1.091\pm0.007$ & 0.222 & $1.076\pm0.006$ & 0.228 \\
5  & 20240309 & $1.152\pm0.014$ & $2.042\pm0.057$ & 0.204 & $2.022\pm0.023$ & $1.842\pm0.086$ & 0.094 & $1.074\pm0.015$ & $1.176\pm0.010$ & 0.097 & $1.216\pm0.009$ & 0.116 \\
6  & 20240310 & $1.104\pm0.014$ & $3.473\pm0.094$ & 0.502 & $2.154\pm0.029$ & $4.119\pm0.207$ & 0.361 & $1.007\pm0.008$ & $1.288\pm0.006$ & 0.066 & $1.291\pm0.005$ & 0.066 \\
7  & 20240311 & $1.024\pm0.005$ & $1.844\pm0.024$ & 0.555 & $1.985\pm0.008$ & $1.768\pm0.034$ & 0.396 & $1.026\pm0.006$ & $1.632\pm0.006$ & 0.254 & $1.649\pm0.005$ & 0.262 \\
8  & 20240312 & $0.961\pm0.003$ & $2.147\pm0.017$ & 3.374 & $1.902\pm0.004$ & $1.906\pm0.021$ & 2.772 & $0.998\pm0.001$ & $1.600\pm0.001$ & 0.084 & $1.598\pm0.001$ & 0.084 \\
9  & 20240313 & $0.938\pm0.006$ & $2.010\pm0.042$ & 0.502 & $1.766\pm0.007$ & $1.349\pm0.034$ & 0.257 & $1.001\pm0.004$ & $1.164\pm0.003$ & 0.081 & $1.164\pm0.002$ & 0.080 \\
10 & 20240315 & $0.991\pm0.033$ & $1.238\pm0.146$ & 0.162 & $2.196\pm0.086$ & $2.439\pm0.347$ & 0.100 & $1.406\pm0.064$ & $0.497\pm0.010$ & 0.156 & $0.542\pm0.014$ & 0.261 \\
11 & 20240317 & $0.977\pm0.011$ & $1.466\pm0.050$ & 0.128 & $1.876\pm0.018$ & $1.215\pm0.063$ & 0.110 & $0.979\pm0.012$ & $0.837\pm0.006$ & 0.081 & $0.831\pm0.005$ & 0.082 \\
12 & 20240321 & $0.978\pm0.015$ & $1.647\pm0.074$ & 0.101 & $1.774\pm0.025$ & $1.092\pm0.090$ & 0.095 & $0.940\pm0.021$ & $0.776\pm0.010$ & 0.113 & $0.764\pm0.009$ & 0.122 \\
13 & 20240327 & $1.220\pm0.010$ & $1.739\pm0.030$ & 0.215 & $2.001\pm0.018$ & $1.377\pm0.053$ & 0.121 & $1.010\pm0.010$ & $1.347\pm0.007$ & 0.116 & $1.353\pm0.006$ & 0.116 \\
14 & 20240330 & $1.869\pm0.011$ & $1.740\pm0.012$ & 0.112 & $5.185\pm0.151$ & $6.902\pm0.324$ & 0.103 & $0.945\pm0.012$ & $1.297\pm0.010$ & 0.189 & $1.247\pm0.008$ & 0.208 \\
15 & 20240401 & $1.442\pm0.011$ & $1.013\pm0.014$ & 0.130 & $2.861\pm0.038$ & $1.700\pm0.065$ & 0.145 & $1.047\pm0.019$ & $1.033\pm0.012$ & 0.282 & $1.054\pm0.009$ & 0.291 \\
16 & 20240403 & $1.256\pm0.015$ & $0.990\pm0.027$ & 0.101 & $2.279\pm0.039$ & $1.106\pm0.071$ & 0.099 & $1.168\pm0.022$ & $0.506\pm0.005$ & 0.106 & $0.536\pm0.007$ & 0.192 \\
17 & 20240421 & $1.114\pm0.060$ & $1.008\pm0.165$ & 0.585 & $2.395\pm0.203$ & $2.808\pm0.591$ & 0.125 & $0.834\pm0.023$ & $0.314\pm0.005$ & 0.059 & $0.299\pm0.006$ & 0.105 \\
18 & 20240425 & $1.339\pm0.019$ & $0.916\pm0.027$ & 0.069 & $2.566\pm0.062$ & $1.284\pm0.101$ & 0.087 & $1.132\pm0.023$ & $0.418\pm0.005$ & 0.079 & $0.436\pm0.006$ & 0.121 \\
19 & 20240430 & $1.095\pm0.015$ & $1.239\pm0.044$ & 0.132 & $1.881\pm0.027$ & $0.917\pm0.064$ & 0.095 & $1.038\pm0.031$ & $0.574\pm0.007$ & 0.292 & $0.583\pm0.009$ & 0.293 \\
20 & 20240505 & $1.173\pm0.022$ & $0.857\pm0.041$ & 0.084 & $1.887\pm0.054$ & $0.541\pm0.085$ & 0.109 & $1.133\pm0.031$ & $0.463\pm0.007$ & 0.142 & $0.482\pm0.007$ & 0.179 \\
21 & 20240512 & $1.555\pm0.038$ & $1.362\pm0.055$ & 0.182 & $3.037\pm0.221$ & $2.343\pm0.430$ & 0.312 & $1.002\pm0.022$ & $0.465\pm0.006$ & 0.094 & $0.465\pm0.005$ & 0.093 \\
22 & 20240714 & $1.189\pm0.027$ & $1.327\pm0.069$ & 0.263 & $2.227\pm0.064$ & $1.572\pm0.166$ & 0.206 & $0.932\pm0.025$ & $0.962\pm0.014$ & 0.135 & $0.939\pm0.012$ & 0.144 \\
23 & 20240721 & $1.112\pm0.018$ & $1.767\pm0.078$ & 0.088 & $2.049\pm0.037$ & $1.652\pm0.141$ & 0.093 & $1.047\pm0.017$ & $0.725\pm0.008$ & 0.074 & $0.737\pm0.007$ & 0.080 \\
24 & 20240723 & $1.113\pm0.029$ & $1.405\pm0.094$ & 0.235 & $1.444\pm0.047$ & $0.301\pm0.098$ & 0.196 & $0.778\pm0.016$ & $0.675\pm0.011$ & 0.056 & $0.607\pm0.012$ & 0.181 \\
25 & 20240725 & $1.217\pm0.007$ & $1.267\pm0.016$ & 0.153 & $2.343\pm0.019$ & $1.609\pm0.047$ & 0.185 & $1.081\pm0.008$ & $0.607\pm0.003$ & 0.144 & $0.624\pm0.003$ & 0.196 \\
26 & 20240727 & $1.179\pm0.015$ & $1.428\pm0.038$ & 0.131 & $2.229\pm0.037$ & $1.617\pm0.102$ & 0.176 & $0.968\pm0.019$ & $0.678\pm0.008$ & 0.171 & $0.669\pm0.006$ & 0.174 \\
27 & 20240728 & $1.121\pm0.016$ & $2.014\pm0.068$ & 0.066 & $2.048\pm0.044$ & $1.942\pm0.174$ & 0.094 & $0.990\pm0.037$ & $0.647\pm0.017$ & 0.221 & $0.644\pm0.012$ & 0.219 \\
28 & 20240729 & $1.237\pm0.008$ & $2.399\pm0.025$ & 0.066 & $2.435\pm0.023$ & $3.472\pm0.096$ & 0.062 & $1.007\pm0.013$ & $0.873\pm0.007$ & 0.111 & $0.875\pm0.006$ & 0.111 \\
29 & 20240731 & $1.203\pm0.016$ & $1.931\pm0.052$ & 0.133 & $2.211\pm0.037$ & $2.206\pm0.130$ & 0.102 & $0.872\pm0.017$ & $0.701\pm0.009$ & 0.140 & $0.673\pm0.008$ & 0.195 \\
30 & 20240803 & $1.216\pm0.016$ & $1.504\pm0.045$ & 0.077 & $1.927\pm0.043$ & $1.074\pm0.109$ & 0.088 & $1.034\pm0.030$ & $0.762\pm0.014$ & 0.170 & $0.773\pm0.010$ & 0.171 \\
31 & 20240806 & $1.053\pm0.061$ & $0.553\pm0.131$ & 0.342 & $2.828\pm0.294$ & $2.502\pm0.629$ & 0.121 & $0.716\pm0.026$ & $0.471\pm0.014$ & 0.051 & $0.397\pm0.012$ & 0.170 \\
32 & 20240809 & $1.347\pm0.024$ & $1.698\pm0.052$ & 0.111 & $2.779\pm0.125$ & $3.447\pm0.363$ & 0.084 & $0.976\pm0.022$ & $0.603\pm0.008$ & 0.082 & $0.599\pm0.007$ & 0.082 \\
33 & 20240815 & $1.187\pm0.036$ & $2.321\pm0.170$ & 0.277 & $1.510\pm0.041$ & $0.644\pm0.143$ & 0.110 & $0.913\pm0.028$ & $0.697\pm0.014$ & 0.116 & $0.676\pm0.010$ & 0.129 \\
34 & 20240818 & $1.135\pm0.033$ & $0.966\pm0.080$ & 0.110 & $2.094\pm0.076$ & $0.921\pm0.162$ & 0.149 & $0.915\pm0.022$ & $0.533\pm0.008$ & 0.063 & $0.520\pm0.007$ & 0.079 \\
35 & 20240822 & $1.050\pm0.038$ & $1.104\pm0.127$ & 0.214 & $1.915\pm0.124$ & $1.172\pm0.341$ & 0.271 & $0.932\pm0.046$ & $0.466\pm0.016$ & 0.169 & $0.451\pm0.011$ & 0.173 \\

\bottomrule
\end{tabular}
\end{adjustbox}
}
\end{adjustbox}
\end{table*}

\bsp
\label{lastpage}
	
\end{document}